\documentclass[aps,twocolumn,10pt, 
superscriptaddress,
footinbib, floatfix,
prx, longbibliography]{revtex4-2}

\usepackage{amsmath, amssymb, bm, braket, bbold, mathtools, graphicx, latexsym, epstopdf, tensor, hyperref, xfrac, comment, float, siunitx, physics, stackengine, bbm, orcidlink, xcolor, tabularx, multirow}
\usepackage[normalem]{ulem} 
\usepackage[ngerman,english]{babel}
\usepackage[utf8]{inputenc}
\usepackage[T1]{fontenc}
\usepackage[nointegrals]{wasysym}
\usepackage[export]{adjustbox}

\newcommand{\tpm}{$\,\pm\,$}

\graphicspath{{figures/}}
\makeatletter
\let\save@mathaccent\mathaccent
\newcommand*\if@single[3]{%
  \setbox0\hbox{${\mathaccent"0362{#1}}^H$}%
  \setbox2\hbox{${\mathaccent"0362{\kern0pt#1}}^H$}%
  \ifdim\ht0=\ht2 #3\else #2\fi
  }
\newcommand*\rel@kern[1]{\kern#1\dimexpr\macc@kerna}
\newcommand*\widebar[1]{\@ifnextchar^{{\wide@bar{#1}{0}}}{\wide@bar{#1}{1}}}
\newcommand*\wide@bar[2]{\if@single{#1}{\wide@bar@{#1}{#2}{1}}{\wide@bar@{#1}{#2}{2}}}
\newcommand*\wide@bar@[3]{%
  \begingroup
  \def\mathaccent##1##2{%
    \let\mathaccent\save@mathaccent
    \if#32 \let\macc@nucleus\first@char \fi
    \setbox\z@\hbox{$\macc@style{\macc@nucleus}_{}$}%
    \setbox\tw@\hbox{$\macc@style{\macc@nucleus}{}_{}$}%
    \dimen@\wd\tw@
    \advance\dimen@-\wd\z@
    \divide\dimen@ 3
    \@tempdima\wd\tw@
    \advance\@tempdima-\scriptspace
    \divide\@tempdima 10
    \advance\dimen@-\@tempdima
    \ifdim\dimen@>\z@ \dimen@0pt\fi
    \rel@kern{0.6}\kern-\dimen@
    \if#31
      \overline{\rel@kern{-0.6}\kern\dimen@\macc@nucleus\rel@kern{0.4}\kern\dimen@}%
      \advance\dimen@0.4\dimexpr\macc@kerna
      \let\final@kern#2%
      \ifdim\dimen@<\z@ \let\final@kern1\fi
      \if\final@kern1 \kern-\dimen@\fi
    \else
      \overline{\rel@kern{-0.6}\kern\dimen@#1}%
    \fi
  }%
  \macc@depth\@ne
  \let\math@bgroup\@empty \let\math@egroup\macc@set@skewchar
  \mathsurround\z@ \frozen@everymath{\mathgroup\macc@group\relax}%
  \macc@set@skewchar\relax
  \let\mathaccentV\macc@nested@a
  \if#31
    \macc@nested@a\relax111{#1}%
  \else
    \def\gobble@till@marker##1\endmarker{}%
    \futurelet\first@char\gobble@till@marker#1\endmarker
    \ifcat\noexpand\first@char A\else
      \def\first@char{}%
    \fi
    \macc@nested@a\relax111{\first@char}%
  \fi
  \endgroup
}
\makeatother

\begin{document}

\title{Detecting high-dimensional entanglement in cold-atom quantum simulators}

\date{\today}

\author{Niklas Euler\,\orcidlink{0009-0009-2401-817X}}
\email[]{euler@physi.uni-heidelberg.de}
\affiliation{Physikalisches Institut, Universit\"at Heidelberg, Im Neuenheimer Feld 226, 69120 Heidelberg, Germany}
\author{Martin G\"{a}rttner\,\orcidlink{0000-0003-1914-7099}}
\email[]{martin.gaerttner@uni-jena.de}
\affiliation{Physikalisches Institut, Universit\"at Heidelberg, Im Neuenheimer Feld 226, 69120 Heidelberg, Germany}
\affiliation{Kirchhoff-Institut f\"{u}r Physik, Universit\"{a}t Heidelberg, Im Neuenheimer Feld 227, 69120 Heidelberg, Germany}
\affiliation{Institut f\"ur Theoretische Physik, Universit\"{a}t Heidelberg, Philosophenweg 16, 69120 Heidelberg, Germany}
\affiliation{Institute of Condensed Matter Theory and Optics, Friedrich-Schiller-University Jena, Max-Wien-Platz 1, 07743 Jena, Germany}

\begin{abstract}

Quantum entanglement has been identified as a crucial concept underlying many intriguing phenomena in condensed matter systems, such as topological phases or many-body localization. 
Recently, instead of considering mere quantifiers of entanglement like entanglement entropy, the study of entanglement structure in terms of the entanglement spectrum has shifted into the focus, leading to new insights into fractional quantum Hall states and topological insulators, among others.
What remains a challenge is the experimental detection of such fine-grained properties of quantum systems. 
The development of protocols for detecting features of the entanglement spectrum in cold-atom systems, which are one of the leading platforms for quantum simulation, is thus highly desirable and will open up new avenues for experimentally exploring quantum many-body physics. Here, we present a method to bound the width of the entanglement spectrum, or entanglement dimension, of cold atoms in lattice geometries, requiring only measurements in two experimentally accessible bases and utilizing ballistic time-of-flight (TOF) expansion. Building on previous proposals for entanglement certification for photon pairs, we first consider entanglement between two atoms of different atomic species and later generalize to higher numbers of atoms per species and multispecies configurations showing multipartite high-dimensional entanglement. Through numerical simulations, we show that our method is robust against typical experimental noise effects and thus will enable high-dimensional entanglement certification in systems of up to eight atoms using currently available experimental techniques.
\end{abstract}

\maketitle  
\section{Introduction}
Since its initial conception inspired by the EPR paradox \cite{Einstein1935}, quantum entanglement  has been identified as a key aspect in the understanding of a plethora of physical phenomena, such as the dynamics of disordered spin systems \cite{Dur2005}, the thermalization of closed quantum systems \cite{Kaufman2016, Horodecki2009}, and even in the context of the black-hole information paradox \cite{Almheiri2013}. In recent years, much attention has been directed toward the effects of entanglement in condensed matter, where it has been linked to topological properties of quantum states \cite{Kitaev2006, Haque2007} and quantum phase transitions \cite{Osterloh2002, Osborne2002, Vidal2003}, among others \cite{Laflorencie2016}. Studying entanglement in these macroscopic systems directly is oftentimes too challenging due to limited experimental control and measurement capabilities.

The development of experimental systems offering quantum control on the level of single particles over recent decades has enabled an alternative approach to studying such genuine quantum phenomena. To simulate complex
quantum systems, one constructs simpler synthetic systems, called quantum simulators, which mimic, or emulate,
the dynamics of the system of interest. In particular, cold atoms trapped in lattice geometries have evolved into the leading platform for quantum simulation of condensed matter systems \cite{Jaksch2000,  Lewenstein2007, Esslinger2010, Bloch2012, Tarruell2018, Hofstetter2018, Altman2021}. Through the application of external fields, model parameters can be tuned within a broad regime ranging from strong repulsive to attractive interactions, equipping the system with an ideal framework to simulate highly entangled quantum states with single-atom-resolved readout \cite{Jaksch1998, Bakr2009, Murmann2015, Schafer2020}. The capability to detect entanglement in these platforms is crucial for the investigation of the aforementioned phenomena, but still faces challenges \cite{Gurvits2004}. Many experimentally available criteria can, in fact, only indicate (``witness") the existence of entanglement in a state qualitatively \cite{Friis2019}.

In this work, we want to go beyond detecting the mere presence of entanglement and instead make statements about the entanglement structure. The standard measure of entanglement for bipartite pure quantum states $\hat{\rho}_{AB}=\ketbra{\psi}$ is the entanglement entropy, defined as $S(\hat{\rho}_A)=S(\hat{\rho}_B)=-\sum_{i=1}^d p_i\,\log{p_i}$, with the reduced density matrix $\hat{\rho}_A = \Tr_B(\hat{\rho}_{AB})$ ($\hat{\rho}_B$ analogously) and its eigenvalues $p_i$ \cite{Horodecki2009}. Even though in many cases much can be learned from this quantity, it contains less information than the full eigenvalue spectrum, also known as the entanglement spectrum, from which it is derived. Therefore, more recently, the entanglement spectrum itself has been used extensively to investigate the role of entanglement in various phenomena, including fractional quantum Hall states \cite{Li2008}, topological insulators and superconductors \cite{Fidkowski2010}, one-dimensional (1D) systems in the scaling regime \cite{Calabrese2008}, emergent irreversibility \cite{Chamon2014, Shaffer2014}, and  many-body localization transitions \cite{Serbyn2016, Geraedts2016}, leading to new insights. Furthermore, the ability to prepare and certify states with a broad entanglement spectrum would enable the execution of quantum algorithms that exploit this property for enhancing run time and robustness \cite{Muthukrishnan2000, Lanyon2009, Neeley2009}.

The number of nonvanishing terms in the entanglement spectrum is known as the entanglement dimension, or Schmidt rank, of the state. It represents the number of terms needed to faithfully represent the quantum state in the product Hilbert space (with generalizations established for mixed states). Standard methods to obtain the entanglement dimension for cold-atom systems available today are based on full state tomography, or on efficient fidelity-measurement schemes, for which the number of required measurement bases scales quadratically, or linearly, respectively, with the local Hilbert-space dimension $L$ \cite{Friis2019}. Recently, advanced methods for accessing information about the entanglement spectrum have been proposed, including Hamiltonian learning \cite{Kokail2021a, Kokail2021b, joshi_exploring_2023}, random measurement schemes \cite{wyderka_probing_2023, liu_characterizing_2023}, and ancillary-system-based readout protocols \cite{pichler_measurement_2016}.  
However, these approaches either make assumptions about the prepared states potentially leading to bias, or pose stringent requirements on experimental capabilities (for a more detailed discussion, see Sec.~\ref{sec:literature}).

We propose an alternative approach to detecting high-dimensional entanglement in systems of lattice-confined ultracold atoms. Our method is inspired by earlier findings for entangled photon pairs in different polarization states \cite{Bavaresco2017}. In that work the authors construct a measurable lower bound on the state fidelity to a highly entangled reference state. This approach provides a powerful tool as one can define a set of fidelity thresholds with each threshold corresponding to a matching minimum entanglement dimension of the measured state \cite{Fickler2014}. Bounds on the fidelity to the reference state thus naturally translate to bounds on the entanglement dimension of the prepared quantum state. One can construct such a bound by measuring in only two mutually unbiased bases (MUB) $\ket{i}_{\mathrm{m}}$ and $\ket{j}_{\mathrm{n}}$, i.e.,\ $\forall \mathrm{m},\mathrm{n}:\tensor[_{\mathrm{m}}]{\braket{i}{j}}{_{\mathrm{n}}} = L^{-1}$, simplifying the experimental procedure significantly. However, implementation of two such MUB measurements for cold-atom systems is a challenging problem.

Our main contribution is to derive lower bounds on the fidelity to highly entangled reference states that only require position- and momentum-correlation measurements, generalizing previously reported bounds in several ways. Both the position and momentum bases can be accessed by measuring the atom positions \textit{in situ} and after TOF expansion \cite{Folling2005, Schafer2020}, techniques that are well established experimentally \cite{Bergschneider2019}. The fidelity bounds directly yield bounds on the entanglement dimension and thus measurable Schmidt-number witnesses. Furthermore, we show that this protocol is applicable to a large class of reference states, to bipartite systems with multiple indistinguishable particles per species (party) for both fermions and hard-core bosons, and even to a multipartite setting. One might expect that a bound based on the fidelity to a reference state gives satisfactory results only for experimental states close to that reference, i.e.,\ for states the reduced density-matrix spectrum of which is similar to that of the reference state. Our findings indicate, however, that our bound detects high-dimensional entanglement for a broad range of quantum states, even in the presence of strong decoherence.
The bound turns out to be robust against typical experimental noise sources and its tightness decreases at most linearly with the noise strength, i.e.,\ with the impurity of the prepared state.

In the remainder of this work, we first establish a fidelity bound for a pair of two entangled atoms in an optical lattice in Sec.~\ref{sec:bounds} and test its robustness regarding typical experimental noise using a Hubbard model in Sec.~\ref{sec:numerics}. Subsequently, we generalize the method to multiple indistinguishable atoms per species (Sec.~\ref{sec:ML}) and to a multipartite setting, where more than two different atomic species are entangled (Sec.~\ref{sec:multipartite}).
In Sec.~\ref{sec:repulsive} we derive fidelity bounds for extended classes of reference states.
Our conclusions and a discussion of our results are provided in Sec.~\ref{sec:conclusions}.

\section{Bound on Entanglement Dimension}
\label{sec:bounds}

\begin{figure*}
    \centering
    \includegraphics[width=\textwidth]{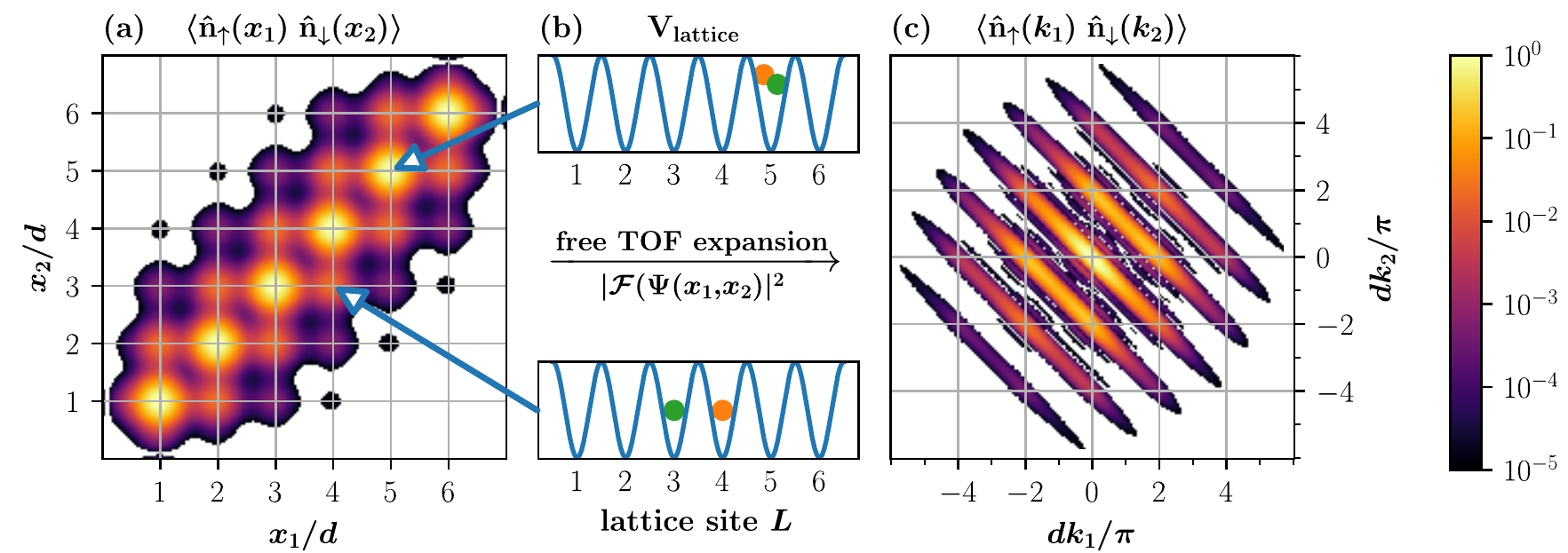}
    \caption{\textbf{(a)} The position-space correlation function $\langle\hat{\mathrm{n}}_{\uparrow}(x_1)\hat{\mathrm{n}}_{\downarrow}(x_2)\rangle$ of the two-particle attractive Hubbard-model ground state for $L=6$ lattice sites with lattice spacing $d$ at $U/J=-12$. \textbf{(b)} A graphical representation of both particles occupying the same lattice site (top) or adjacent lattice sites (bottom) with the respective signals in (a). \textbf{(c)} The momentum correlation function $\langle\hat{\mathrm{n}}_{\uparrow}(k_1)\hat{\mathrm{n}}_{\downarrow}(k_2)\rangle$ corresponding to the position correlation function of (a). All values smaller than \num{1e-5} in both (a) and (c) have been masked.}
    \label{fig:fourier-trafo}
\end{figure*}

Any bipartite pure quantum state on a product Hilbert space $\mathcal{H} = \mathcal{H}_A\otimes\mathcal{H}_B$ can be represented in Schmidt-decomposed form $\ket{\psi}_{\mathrm{AB}} = \sum_{i=1}^{k}\lambda_i\, \ket{i}_\mathrm{A}\otimes\ket{i}_{\mathrm{B}}$, a basis choice that minimizes the number of contributions $k$ (also known as Schmidt rank or entanglement dimension, $D_{\mathrm{ent}}$) needed to represent a given quantum state \cite{Schmidt1907}. Any separable state can be written through one tensor-product term alone and therefore $k=1$. The defining feature of entangled states is that this no longer holds true and thus $k\geq2$, as can be seen on the example of the singlet state $\ket{\Psi}_{\mathrm{EPR}}=(\ket{\uparrow\downarrow}-\ket{\downarrow\uparrow})/\sqrt{2}$. The composition of these tensor-product contributions and their weights defines the entanglement structure of a quantum state. Gaining information on the full structure is an exceedingly hard problem for multidimensional states, i.e.,\ states entangled in several internal degrees of freedom, or multipartite states, i.e.,\ states made up of three or more entangled parties. Determining the entanglement dimension instead is both insightful and experimentally feasible, as we show in this paper. We start with the case of attractive interactions and later, in Sec.~\ref{sec:repulsive}, we generalize to the repulsive case.

The entanglement dimension of a bipartite system is bounded by the size of the smaller of the two local Hilbert spaces,  $k_{\mathrm{max}} = \min[\mathrm{dim}(\mathcal{H}_A), \, \mathrm{dim}(\mathcal{H}_B)] \eqqcolon L$. In the remainder of this work we take the local Hilbert-space dimensions to be equal and finite.
One can choose a maximally entangled state (MES) of the system that has equal coefficients for all $L$ terms,
\begin{equation}
    \ket{\Psi}_{\mathrm{MES}} = \frac{1}{\sqrt{L}}\sum_{{m=1}}^{L}\ket{mm}.\label{eq:MES}
\end{equation}
This highly entangled state acts as a reference to which experimentally prepared states $\hat{\rho}$ can be compared. To give some intuition, the two subsystems will later be the two atoms in the lattice, where $m$ labels the lattice sites. The fidelity of the experimental state $\hat{\rho}$ to the reference MES,
\begin{equation}
    F(\hat{\rho}, \Psi_{\mathrm{MES}}) = \frac{1}{L}\sum\limits_{\mathclap{m,n=1}}^{L}\bra{mm}\hat{\rho}\ket{nn},\label{eq:fidel}
\end{equation}
implies a convenient state-distance measure to compare the two states, as it is bounded as a function of the entanglement dimension of $\hat{\rho}$. One can explicitly construct a set of bounds B$_k$ on the fidelity to the MES, $F(\hat{\rho}, \Psi_{\mathrm{MES}})$, given by
\begin{equation}
   F(\hat{\rho}, \Psi_{\mathrm{MES}}) \leq \mathrm{B}_k(\Psi_{\mathrm{MES}}) = \frac{k}{L},\label{eq:Bk}
\end{equation}
which hold for any experimental state $\hat{\rho}$ with entanglement dimension $D_\mathrm{ent}\leq k$ \cite{Piani2007,Fickler2014}.
In the case of mixed states, $\hat{\rho}=\sum_i p_i\ketbra{\psi_i}$, the notion of an entanglement dimension has to be extended to the so-called Schmidt number. This is defined as the maximum entanglement dimension of the pure parts $\ket{\psi_i}$ of the state, minimized over all possible pure-state decompositions: $D_\mathrm{ent}(\hat{\rho})=\min_{\mathrm{decomp.}}\{\max_i[D_\mathrm{ent}(\psi_i)]\}$ \cite{Horodecki2009}. The violation of the relation in Eq.~\eqref{eq:Bk} for given $k$ therefore indicates that $\hat{\rho}$ is entangled with a dimension of $k + 1$ or higher.
This not only gives a robust entanglement witness, since the lowest threshold $\mathrm{B}_1=L^{-1}$ already indicates entanglement, but also bounds the width of the entanglement spectrum and hence gives insight into the entanglement structure.
We exemplify this in Sec.~\ref{sec:lattice} for localized dimer states in a Hubbard model where the detected entanglement dimension correlates with the number of macroscopic Schmidt coefficients. Additionally, one can use the fidelity to construct lower bounds on the entanglement of formation, as has been shown in Refs.~\cite{Huber2013a, Bavaresco2017}, establishing $F(\hat{\rho}, \Psi_{\mathrm{MES}})$ as a versatile source of information about the entanglement content of $\hat{\rho}$.
Nonetheless, fidelity measurements come with significant experimental complexity, in general requiring measuring in $L+1$ different bases for an $L$-dimensional local Hilbert space \cite{Bavaresco2017}. 

In the following, we establish a lower bound $\tilde{F}(\hat{\rho}, \Psi_{\mathrm{MES}})$ on the fidelity $F(\hat{\rho}, \Psi_{\mathrm{MES}})$ accessible to experiments with cold atoms in optical lattices (or arrays of optical tweezers). It only requires measurements in two bases, independent of the local Hilbert-space dimension given by the lattice size. We start with the case of two distinguishable atoms here and develop generalizations to higher atom numbers and other reference states in later sections. The two atoms constitute the two entangled parties and their local Hilbert spaces are spanned by the discrete position states (sites) of the atoms in the optical-lattice potential. Consequently, the MES for this product basis is a superposition state with both atoms located at the same lattice site, in a superposition summing over all $L$ sites with equal probability [Eq.~\eqref{eq:MES}]. 

It is insightful to split the fidelity into two sums,
\begin{equation}
F(\hat{\rho}, \Psi_{\mathrm{MES}}) = \sum\limits_{m=1}^L\frac{\bra{mm}\hat{\rho}\ket{mm}}{L} + \underbrace{\sum\limits_{\mathclap{\substack{m,n = 1 \\ m\neq n}}}^L \frac{\bra{mm}\hat{\rho}\ket{nn}}{L}}_{F_{\mathrm{coh}}},\label{eq:fidelsplit}
\end{equation}
dividing the contributions into state populations (left-hand sum) and two-particle coherences $F_{\mathrm{coh}}$ (right-hand sum). The state populations of the two distinguishable species, in the following labeled as $\uparrow$ and $\downarrow$ with their corresponding number operators $\hat{\mathrm{n}}_{\uparrow}$ and $\hat{\mathrm{n}}_{\downarrow}$, can be obtained by spatially discretizing the joint density distribution $\langle\hat{\mathrm{n}}_{\uparrow}(x_1)\hat{\mathrm{n}}_{\downarrow}(x_2)\rangle$. It can be probed directly through single particle resolved fluorescence imaging, realizing high-precision \textit{in situ} measurements \cite{Bergschneider2019, Schafer2020, Bonneau2018}. A representation of $\langle\hat{\mathrm{n}}_{\uparrow}(x_1)\hat{\mathrm{n}}_{\downarrow}(x_2)\rangle$ for the ground state of a Hubbard Hamiltonian with $L = 6$ at $U/J = -12$ is displayed in Fig.~\ref{fig:fourier-trafo}(a) (for details on the model and numerical implementation, see Sec.~\ref{sec:numerics}). Each grid point represents a two-particle state contributing to $\hat{\rho}$. The signals on the diagonal represent dimer population probabilities, whereas off-diagonal elements correspond to configurations with atoms on different sites [Fig.~\ref{fig:fourier-trafo}(b)]. The wave-function envelope is determined by the on-site Wannier basis of the lattice and depends on the lattice depth $V_0$ and site spacing $d$. Since $\sum_{m=1}^L\bra{mm}\hat{\rho}\ket{mm}/L\leq1/L$, it is clear that the populations contribute at most $\propto1/L$ to $F$. Their impact therefore becomes negligible compared to coherences for large systems.

Such direct experimental access is not available for the two-particle coherences $F_{\mathrm{coh}}$, but one can instead bound $F_{\mathrm{coh}}$ from below by measuring in a second basis. A natural choice for cold atoms is the momentum basis, as the system comes with a native implementation of the corresponding basis change, the Fourier transformation. It can be applied efficiently by rapidly switching off the lattice potential and interactions and subsequently letting the atoms propagate in a weak harmonic potential for $t = T/4$ with trap oscillation period $T$ before taking a fluorescence image \cite{Bergschneider2019, brown_angle-resolved_2020, Schafer2020}. By repeatedly preparing and measuring a state with this scheme, one acquires samples from the momentum correlation function $\langle\hat{\mathrm{n}}_{\uparrow}(k_1)\hat{\mathrm{n}}_{\downarrow}(k_2)\rangle$. To show how to utilize this to bound state coherences from below, we construct the corresponding measurement operator by stating the effect of the Fourier transform on the localized Wannier basis functions of the lattice potential. The basis function for the $n$th lattice site can be expressed as $\omega(x-nd)$ due to the discrete translational invariance of the lattice, where $d$ is the lattice spacing. Any shift in position space causes a phase factor in momentum space, so one obtains
\begin{equation}
    \mathcal{F}[w(x-nd)](k) = \tilde{\omega}(k)\exp(indk)
\end{equation}
for the single-atom wave function in momentum space with $\tilde{\omega}(k)$ being the Fourier transform of the Wannier envelope. \cite{Bergschneider2019}. Using the field operators,
\begin{align}
\begin{split}
        \hat{\Psi}^{\dagger}_{\sigma}(k) &= \tilde{\omega}(k)^*\sum_{j=1}^L e^{-idkj}\hat{c}^{\dagger}_{j,\sigma},\\
        \hat{\Psi}_{\sigma}(k) &= \tilde{\omega}(k)\sum_{j=1}^L e^{idkj}\hat{c}_{j,\sigma},
\end{split}
\end{align}
defined via lattice-site creation (annihilation) operators $\hat{c}^{\dagger}_{j,\sigma}$ ($\hat{c}_{j,\sigma}$), one can represent the particle number operator in momentum space in the position-space basis \hbox{$\{\ket{j}\,|\,j\in\{1,\,...\,L\}\}$} as 
\begin{equation}
    \hat{\mathrm{n}}_{\sigma}(k) =\hat{\Psi}^\dagger_{\sigma}(k)\hat{\Psi}_{\sigma}(k) = |\tilde{\omega}(k)|^2\sum_{m,n = 1}^L\ketbra{m}{n}e^{id(m-n)k}.
\end{equation}
As only one particle per species is present in the lattice, no differentiation between fermions and bosons has to be made here.
The full expectation value $\langle\hat{\mathrm{n}}_{\uparrow}(k_1)\hat{\mathrm{n}}_{\downarrow}(k_2)\rangle$ in the density-matrix picture is given by the trace over the product of the two momentum number operators and the density matrix, $\langle\hat{\mathrm{n}}_{\uparrow}(k_1)\hat{\mathrm{n}}_{\downarrow}(k_2)\rangle = \Tr{\hat{\mathrm{n}}_{\uparrow}(k_1)\hat{\mathrm{n}}_{\downarrow}(k_2)\hat{\rho}}$. Finally, by exploiting the cyclic property of the trace, one arrives at the following expression:

\begin{subequations}
    \label{eq:pos_space_rep}
    \begin{align}
        \langle\hat{\mathrm{n}}_{\uparrow}(k_1)\hat{\mathrm{n}}_{\downarrow}(k_2)\rangle &= \sum\limits_{\mathclap{m,n,m',n' \,=\, 1}}^L \phi_{mnm'n'}(k_1,k_2)\bra{mn}\hat{\rho}\ket{m'n'},\label{eq:momcorr_phi}\\
        \phi_{mnm'n'}(k_1,k_2) &= |\tilde{w}(k_1,k_2)|^2 e^{-id[(m-m')k_1 + (n-n')k_2]}.\label{eq:phi}
    \end{align}
\end{subequations}

Each density-matrix element $\bra{mn}\hat{\rho}\ket{m'n'}$ is weighted by $\phi_{mnm'n'}(k_1,k_2)$ [Eq.~\eqref{eq:phi}], containing the Fourier-transformed Wannier envelope $\tilde{\omega}(k_1,k_2) \coloneqq \tilde{\omega}(k_1)\tilde{\omega}(k_2)$ and a phase factor obtained through the Fourier transformation \cite{Bonneau2018}.  The above-given description is naturally rewritten in terms of a new set of basis functions  $\{\varphi^{\mathrm{R}}_{\alpha\beta},\varphi^{\mathrm{I}}_{\alpha\beta}\}$,
\begin{subequations}
    \begin{align}
    \varphi^{\mathrm{R}}_{\alpha\beta}=|\tilde{\omega}(k_1,k_2)|^2\cos[d(\alpha k_1+\beta k_2)],\\
    \varphi^{\mathrm{I}}_{\alpha\beta}=|\tilde{\omega}(k_1,k_2)|^2\sin[d(\alpha k_1+\beta k_2)],
    \end{align}
    \label{eq:basis}
\end{subequations}
by bundling terms with the same complex phase factors and their conjugate counterparts into trigonometric basis functions of the two lattice momenta $k_1$ and $k_2$. The full momentum correlation function then reads
\begin{widetext}
\begin{subequations}\label{eq:grpdecomposition}
     \begin{gather}
        \langle\hat{\mathrm{n}}_{\uparrow}(k_1)\hat{\mathrm{n}}_{\downarrow}(k_2)\rangle = \sum\limits_{{\substack{(\alpha,\beta)\in M}}}\operatorname{Re}(g_{\alpha\beta})\varphi^{\mathrm{R}}_{\alpha\beta} - \operatorname{Im}(g_{\alpha\beta})\varphi^{\mathrm{I}}_{\alpha\beta},\label{eq:decomposition}\\
        g_{\alpha\beta} = 2\sum\limits_{m,n=1}^{L}\bra{mn}\hat{\rho}\ket{(m+\alpha), (n+\beta)} \quad \mathrm{with}\quad m+\alpha, n+\beta\in\{1\ldots L\},\quad g_{00}=1,\label{eq:coeffs}\\
        M = \left\{(\alpha,\beta) \in \{0,\ldots,L-1\}\times\{-(L-1),\ldots,L-1\}~|~ \alpha\neq0\lor\beta\geq0\right\}.\label{eq:M}
    \end{gather}
\end{subequations}
\end{widetext}
The above-mentioned basis weights $\operatorname{Re}(g_{\alpha\beta})$ and $\operatorname{Im}(g_{\alpha\beta})$ in Eq.~\eqref{eq:decomposition} are sums over the real and imaginary parts of the coherences of the density matrix $\hat{\rho}$ [see Eq.~\eqref{eq:coeffs}]. Each coefficient $g_{\alpha\beta}$ is defined by the pair of position-space distances for all contributing coherences $\bra{mn}\hat{\rho}\ket{m'n'}$ to $g_{\alpha\beta}$ with $(\alpha,\beta)=(m-m',n-n')$.
The set of all coherences contributing to a given coefficient can simply be constructed by shifting all atom positions of one of the coherences along the lattice. Since we have already combined coherences and the corresponding phase factors with their complex conjugates, we have to introduce the index set $M$ in Eq.~\eqref{eq:M} to avoid double counting of coherences. For additional information regarding Eq.~\eqref{eq:grpdecomposition}, we refer the reader to Ref.~\cite{Bergschneider2019}. 
Obtaining the coefficients $g_{\alpha\beta}$ is not directly straightforward, as the basis $\{\varphi^{\mathrm{R}}_{\alpha\beta},\varphi^{\mathrm{I}}_{\alpha\beta}\}$ is nonorthogonal due to the modulation of the periodic basis functions through the envelope $|\tilde{w}(k_1,k_2)|^2$ \cite{Bergschneider2019}. Projecting the measured distribution [cf.\ Eq.~\eqref{eq:decomposition}] onto the basis function set therefore yields smeared-out coefficients $c_{\alpha\beta}$, 
 \begin{subequations}
    \label{eq:projectionInt}
     \begin{align}
        \begin{split}
            \operatorname{Re}(c_{\alpha\beta})&=\int\mathrm{d}k_1\mathrm{d}k_2\,\langle\hat{\mathrm{n}}_{\uparrow}(k_1)\hat{\mathrm{n}}_{\downarrow}(k_2)\rangle \varphi^{\mathrm{R}}_{\alpha\beta},\label{eq:cosproj}
        \end{split}\\
        \begin{split}
            \operatorname{Im}(c_{\alpha\beta})&=\int\mathrm{d}k_1\mathrm{d}k_2\,\langle\hat{\mathrm{n}}_{\uparrow}(k_1)\hat{\mathrm{n}}_{\downarrow}(k_2)\rangle\varphi^{\mathrm{I}}_{\alpha\beta},\label{eq:sinproj}
        \end{split}
    \end{align}
\end{subequations}
where each coefficient also contains small contributions coming from the nonvanishing overlap with other basis elements. To overcome this problem, we explicitly construct the linear transformation $\bm{Q}$ that maps the set of actual basis weights $\Vec{G}$ to the measured coefficients $c_{\alpha\beta}$ contained in $\Vec{C}$,
\begin{equation}
    \Vec{C} = \bm{Q}\Vec{G},\label{eq:nonorthogonal}
\end{equation}
where each element of the matrix $\bm{Q}$ is given by an overlap integral between a pair of basis functions (for details, see Appendix \ref{app:numerics}). These integrals factorize since the Fourier-transformed Wannier envelope factorizes as well; consequently, only a small number of 1D integrals linear in the number of lattice sites must be computed to construct $\bm{Q}$. The actual basis weights $g_{\alpha\beta}$ are then extracted by formally inverting $\bm{Q}$ and rewriting Eq.~\eqref{eq:nonorthogonal} as 
\begin{equation}
    \Vec{G} = \bm{Q}^{-1}\Vec{C}\label{eq:lintrafo}.
\end{equation}
Numerically, we employ a conjugate-gradient method to determine $ \Vec{G}$. The two projection integrals in Eqs.~\eqref{eq:cosproj} and \eqref{eq:sinproj} can be evaluated in a simplified way using Monte Carlo importance-sampling techniques. By treating the momentum correlation function as a normalizable multivariate probability density, it can be absorbed in a redefinition of the integration variable. The remaining integrals,
 \begin{subequations}
    \label{eq:projectionIntMC}
     \begin{align}
        \begin{split}
            \operatorname{Re}(c_{\alpha\beta}) = \langle\varphi^{\mathrm{R}}_{\alpha\beta}\rangle_{k_1,k_2\sim\langle\hat{\mathrm{n}}_{\uparrow}(k_1)\hat{\mathrm{n}}_{\downarrow}(k_2)\rangle},\label{eq:cosprojMC}
        \end{split}\\
        \begin{split}
            \operatorname{Im}(c_{\alpha\beta}) = \langle\varphi^{\mathrm{I}}_{\alpha\beta}\rangle_{k_1,k_2\sim\langle\hat{\mathrm{n}}_{\uparrow}(k_1)\hat{\mathrm{n}}_{\downarrow}(k_2)\rangle},\label{eq:sinprojMC}
        \end{split}
    \end{align}
\end{subequations}
are then directly evaluated through the measured or simulated momentum correlation samples. This evaluation method enables scalability to higher atom numbers introduced later, as the Monte Carlo integration error scaling is independent of the integral dimension, while also reducing the variance of the integrand at the same time. We make some additional comments regarding synthetic data generation and efficient computation of $\bm{Q}$ in Appendix~\ref{app:numerics}. 

At this point one has obtained access to basis weights $g_{\alpha\beta}$ equal to sums over subsets of coherences of $\hat{\rho}$. However, not only the two-particle coherences relevant for $F_{\mathrm{coh}}$ in Eq.~\eqref{eq:fidelsplit} are contained within the basis weights $g_{\alpha \beta}$ but also different-site two-particle coherences that do not contribute to the fidelity $F(\hat{\rho}, \Psi_{\mathrm{MES}})$. We note that any general density-matrix element is bounded from above by using Cauchy-Schwarz inequality
\begin{align}
    \begin{split}
        \operatorname{Re}(\bra{mn}\hat{\rho}\ket{m'n'})&\, \leq |\bra{mn}\hat{\rho}\ket{m'n'}|\\
        &\overset{\mathrm{CSI}}{\leq} \sqrt{\bra{m'n'}\hat{\rho}\ket{m'n'}\bra{mn}\hat{\rho}\ket{mn}},\label{eq:CSI}
    \end{split}
\end{align}
with the right-hand side containing only already measured state populations and thus adding no new experimental complexity. For pure states, the second inequality in Eq.~\eqref{eq:CSI} is obviously tight but it grows looser with increasing mixedness of the state. In the two-atom case presented here, the subset of $(\alpha,\,\beta)\in M$ that carries relevant two-particle coherences reduces to \hbox{$\alpha=\beta\eqqcolon \delta \in \{1,\ldots,\,L-1\}$}. The desired sum of relevant coherences can then be lower bounded by subtracting the bounds in Eq.~\eqref{eq:CSI} for all noncontributing coherences from the sum of relevant basis coefficients,
\begin{align}
    \begin{split}
        &\sum\limits_{\mathclap{\substack{m,n = 1 \\ m\neq n}}}^L \frac{\bra{mm}\hat{\rho}\ket{nn}}{L} = \sum\limits_{\mathclap{\substack{m,n = 1 \\ m < n}}}^L \frac{2\operatorname{Re}(\bra{mm}\hat{\rho}\ket{nn})}{L} \geq\\
        &\sum\limits_{\mathclap{\delta=1}}^{L-1}\left(\frac{\operatorname{Re}(g_{\delta\delta})}{L}-2\sum\limits_{\mathclap{\substack{m,n = 1\\m\neq n}}}^{L-\delta} \frac{\sqrt{\bra{m'n'}\hat{\rho}\ket{m'n'} \bra{mn}\hat{\rho}\ket{mn}}}{L}\right)\\
        &\hspace{2.55cm}\eqqcolon\, \tilde{F}_{\mathrm{coh}}(\hat{\rho}, \Psi_{\mathrm{MES}})\,\label{eq:fidelbound}\\
        &\hspace{2.55cm}\mathrm{with}\quad m' \coloneqq m + \delta, \quad n' \coloneqq n + \delta
    \end{split}
\end{align}
where the second sum of the last expression covers all nondimer coherences. Together with the same-site populations displayed in the first sum of Eq.~\eqref{eq:fidelsplit}, we formulate the complete experimentally accessible lower bound on the fidelity of the experimental state $\hat{\rho}$ to $\Psi_{\mathrm{MES}}$ as
\begin{equation}
    \tilde{F}(\hat{\rho}, \Psi_{\mathrm{MES}}) = \sum\limits_{m=1}^L\frac{\bra{mm}\hat{\rho}\ket{mm}}{L} + \tilde{F}_{\mathrm{coh}}(\hat{\rho}, \Psi_{\mathrm{MES}})\,.
    \label{eq:fidelbound_full}
\end{equation}
Inserting this bound in Eq.~\eqref{eq:Bk} yields our first main result,
\begin{equation}
    \tilde{F}(\hat{\rho}, \Psi_{\mathrm{MES}}) \leq F(\hat{\rho}, \Psi_{\mathrm{MES}}) \leq \mathrm{B}_k(\Psi_{\mathrm{MES}})\,,\label{eq:inequal}
\end{equation}
where the fidelity bound $\tilde{F}$ constitutes an entanglement-dimension witness and is obtainable directly through fluorescence measurements, \textit{in situ} and after TOF. Thus, if $\tilde{F}(\hat{\rho}, \Psi_{\mathrm{MES}})$ exceeds $\mathrm{B}_k(\Psi_{\mathrm{MES}})$ for some given $k$, the state $\hat{\rho}$ is certified to be entangled in at least $k+1$ dimensions.

The proposed experimental protocol can be summarized as follows: 
One prepares an ensemble of two atoms of different species in a periodic potential in some state of interest.
\begin{enumerate}
    \item By single-atom-resolved detection, one measures the position-space correlation function $\langle\hat{\mathrm{n}}_{\uparrow}(x_1)\hat{\mathrm{n}}_{\downarrow}(x_2)\rangle$. The signal is discretized by identifying the atom positions obtained in each shot with a pair of lattice sites, which yields the position-space populations $\bra{mn}\hat{\rho}\ket{mn}$ entering in Eqs.~\eqref{eq:fidelbound} and \eqref{eq:fidelbound_full}.
    \item The momentum-space distribution is probed through ballistic TOF expansion, resulting in an effective Fourier transformation of the wave function. The coefficients $c_{\alpha\beta}$ are obtained from the measured momentum correlation function, $\langle\hat{\mathrm{n}}_{\uparrow}(k_1)\hat{\mathrm{n}}_{\downarrow}(k_2)\rangle$, by computing the overlap between the measured distribution and the trigonometric basis functions $\{\varphi^{\mathrm{R}}_{\alpha\beta},\varphi^{\mathrm{I}}_{\alpha\beta}\}$ [Eq.~\eqref{eq:basis}], i.e.,\ by evaluating the basis functions using the sampled momenta. From these, the corrected expansion coefficients $g_{\alpha\beta}$ are obtained via Eq.~\eqref{eq:lintrafo} and inserted into Eq.~\eqref{eq:fidelbound}, which yields the desired lower bound on the reference-state fidelity in Eq.~\eqref{eq:fidelbound_full}.
\end{enumerate}
The statistical requirements for confident certification are discussed in Sec.~\ref{sec:statistics} and a study of the robustness of the protocol with respect to typical experimental noise effects is given in 
Secs.~\ref{sec:dephasing} and \ref{sec:lattice}.

\section{Certification Robustness Under Realistic Conditions}
\label{sec:numerics}

We study the performance of our method under realistic experimental conditions through numerical simulations. Our model system is a 1D open-boundary Hubbard model, realized by cold atoms in a deep optical lattice ($V_0 = 8E_r$ \footnote{recoil energy $E_r =\frac{\hbar^2\pi^2}{2md^2}$}) in the tight-binding approximation \cite{Wall2015}. Due to limited wave-function overlap between sites, all tunneling going beyond adjacent sites is suppressed. Through the application of external magnetic fields, Feshbach resonances can be utilized to implement an effective on-site atom-atom interaction with a highly tuneable interaction strength \cite{Feshbach1958}. The dynamics of the system are captured by the Hamiltonian
\begin{equation}
    \hat{H} = -J\sum\limits_{\sigma}\sum\limits_{i}(\hat{c}_{i,\sigma}^\dagger \hat{c}_{i+1,\sigma}^{\phantom{\dagger}} + \mathrm{h.c.}) + U\sum\limits_{i}\hat{\mathrm{n}}_{i\downarrow}\hat{\mathrm{n}}_{i\uparrow}\,,\label{eq:Hamiltonian}
\end{equation}
with the tunneling strength $J$, interaction strength $U$, creation (annihilation) operator $\hat{c}_{i,\sigma}^\dagger$ ($\hat{c}_{i,\sigma}$) for an atom on site $i$ and in spin state $\sigma \in \left\{\uparrow,\,\downarrow\right\}$, and their corresponding atom number operators $\hat{\mathrm{n}}_{i\downarrow},\,\hat{\mathrm{n}}_{i\uparrow}$ \cite{Hubbard1963}. This system is characterized by the ratio $U/J$ ($J>0$), where negative values correspond to attractive and positive values to repulsive interactions. In the simple case of two distinguishable particles, both Fermi-Dirac and Bose-Einstein statistics produce the same dynamics. The Hubbard model was chosen due to its simplicity and widespread use in numerical modeling \cite{Wall2015, Tarruell2018} but our readout scheme is also applicable to other lattice Hamiltonians.

\begin{figure}
    \centering
    \includegraphics[width=\columnwidth]{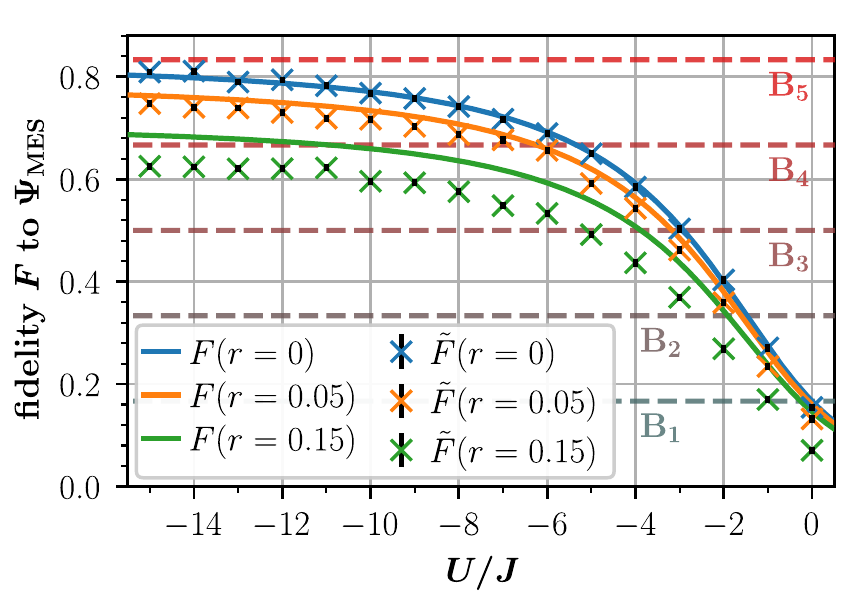}
    \caption{The dependence of the fidelity $F$ and the fidelity bound $\tilde{F}$ on the interaction-to-tunneling-strength ratio $U/J$ for pure ($r = 0$) and dephased ($r \in \{0.05,~0.15\}$) ground states. The B$_k$ thresholds are the horizontal dashed lines such that fidelities above any B$_k$ indicate at least $k+1$ entanglement dimensions. Both $F$ and $\tilde{F}$ increase with growing attractive interaction strength before saturation. The tightness of $\tilde{F}$ decreases with increasing mixing rate $r$. The statistical error bars are small and barely visible.}
    \label{fig:fidelity}
\end{figure}

\begin{figure}
    \centering
    \includegraphics[width=\columnwidth]{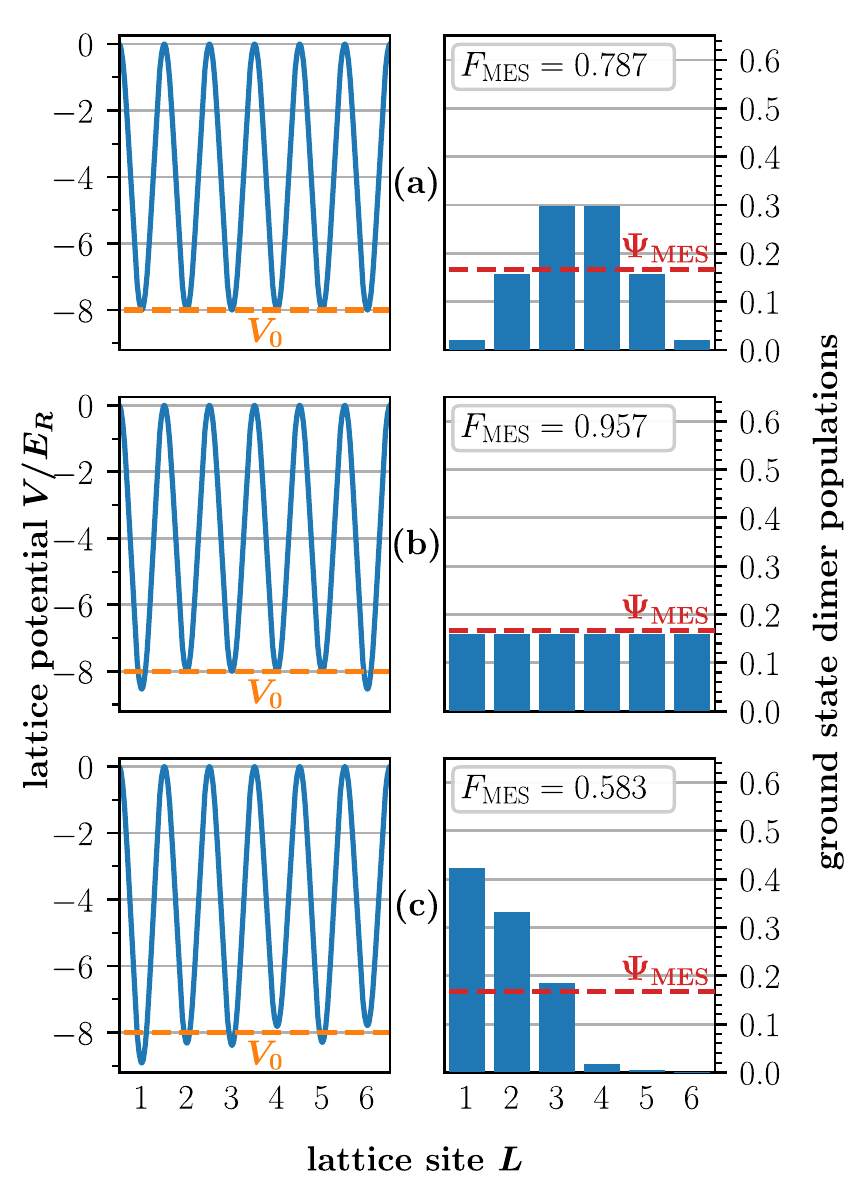}
    \caption{The lattice-potential configurations in units of the recoil energy $E_R$ (left) and the respective dimer-occupation probability distributions for the ground state (right). The dashed lines indicate the potential baseline depth $V_0$ (orange) and the uniform probability distribution of the MES $\Psi_{\mathrm{MES}}$ (red). In all three cases, the ground state has the maximum entanglement dimension of $D_\mathrm{ent}=6$.  \textbf{(a)} An even unaltered lattice potential. The dimer population is heavily centered on the central lattice sites. One can certify up to $D_{\mathrm{ent}}=5$. \textbf{(b)} A lattice with increased potential depth at the outlying sites, resulting in a uniform distribution among all lattice sites and full certification of $D_{\mathrm{ent}}=6$. \textbf{(c)} A lattice with potential fluctuations $\Delta_E\sim\mathcal{N}(0,\, 0.08J)$ on each lattice site. The dimer population shows strong localization and is far away from $\Psi_{\mathrm{MES}}$. Nevertheless, $D_{\mathrm{ent}}=4$ can be certified.}
    \label{fig:lattice-config}
\end{figure}

In the remainder of this section, we consider the ground state of the two atoms in a lattice of size $L=6$ with attractive interactions at $U/J = -12$ and use \num{2.5e4} momentum-space and \num{1e4} position-space samples for certification, unless specified otherwise. Later, in Sec.~\ref{sec:repulsive}, we will also consider repulsive interactions, where robust entanglement certification is achieved by adapting the employed reference state. In the configuration given above, $\hat{\rho}$ is entangled in all six lattice degrees of freedom, meaning that $D_{\mathrm{ent}} = 6$, and thus serves as a suitable test state for our entanglement-detection scheme. Figure~\ref{fig:fourier-trafo}(a) shows a representation of the position-space probability distribution. Through exact diagonalization, we find that the fidelity $F(\hat{\rho}, \Psi_{\mathrm{MES}})$ increases with growing attractive interaction strength (blue line in Fig.~\ref{fig:fidelity}) but the ground state does not converge to $\Psi_{\mathrm{MES}}$ ($F(\hat{\rho}, \Psi_{\mathrm{MES}}) < 1$). Knowledge of the exact fidelity would enable us to certify five out of the six entanglement dimensions for moderately attractive interactions [$\tilde{F}(\hat{\rho}, \Psi_{\mathrm{MES}}) > \mathrm{B}_4~\mathrm{for}~U/J \lesssim -6$]. The offset in fidelity with the MES is an effect of the finite system size, as central sites are energetically favored for open boundary conditions, since more tunneling pathways are available [see Fig.~\ref{fig:lattice-config}(a)], making the distribution of populations nonuniform. Since the Schmidt coefficients are given by the double-occupation probabilities in the strongly attractive limit, this behavior translates to a nonuniform entanglement spectrum.
In the case of the pure ground state, we find our fidelity bound to be tight (blue markers in Fig.~\ref{fig:fidelity}). The use of our protocol therefore yields the highest certifiable entanglement dimension achievable using the fidelity to the MES.

In the following subsections, we discuss the requirements of our detection scheme in terms of its robustness with respect to typical experimental imperfections and noise sources, starting with finite measurement statistics in Sec.~\ref{sec:statistics}. In Sec.~\ref{sec:dephasing} we study the effect of generic dephasing noise arising in experiments, which can be caused by, e.g., fluctuations of trapping light parameters or control fields during state preparation. Finally, in Sec.~\ref{sec:lattice}, we consider fluctuations of the depth of individual lattice sites that are characteristic of realizations using arrays of optical tweezers. Randomized potentials lead to localization of atom pairs and thus potentially to a reduction of the entanglement dimension, an effect that becomes observable through our detection scheme. Moreover, in Sec.~\ref{sec:scalability}, we discuss the behavior of our approach in the limit of large lattice sizes. Additionally, in Appendix~\ref{app:thermal}, we have simulated the performance of the bound on thermal ensembles . Readers more interested in generalizations of the scheme may jump to the last paragraph of Sec.~\ref{sec:numerics}, where our results on noise robustness are summarized.

For the simulation results presented in Fig.~\ref{fig:fidelity}, the first two effects are already addressed within the simulation. Finite measurement statistics induce fluctuations of the certified fidelity and thus impact entanglement detection. Additionally, experimental quantum state realizations are in general not pure wave functions $\ket{\psi_0}$ but face mixing and decoherence. A simple model to account for this is to replace $\ket{\psi_0}$ with a dephased density matrix $\hat{\rho} = (1-r) \ketbra{\psi_0} + rL^{-2}\mathbbm{1}$, with a mixing parameter $r$ related to the state impurity $\bar{p} = 1-p$, washing out the probability distribution.
Both effects have been included to produce dephased ground states in Fig.~\ref{fig:fidelity}, each with sampled correlation functions at mixing strengths $r = 0.05$ ($\bar{p}\approx0.095$) and $r = 0.15$ ($\bar{p}\approx0.270$), respectively (orange and green data sets throughout this work). These model alterations clearly lead to loss of bound tightness and add random noise to the fidelity bound $\tilde{F}$, as visible in Fig.~\ref{fig:fidelity}. We discuss these matters in more detail in the following.

\subsection{Sampling statistics}
\label{sec:statistics}

In experiments, both the joint position-space distribution $\langle\hat{\mathrm{n}}_{\uparrow}(x_1)\hat{\mathrm{n}}_{\downarrow}(x_2)\rangle$ and the momentum-space distribution $\langle\hat{\mathrm{n}}_{\uparrow}(k_1)\hat{\mathrm{n}}_{\downarrow}(k_2)\rangle$ are probed by repeated state preparation and measurement, each experimental run providing one sample point drawn from the respective distribution. The finite sample numbers are the cause of statistical errors in our fidelity-bound estimation. In this section, we systematically explore the scaling of the standard error (SE) of our bound $\tilde{F}(\hat{\rho}, \Psi_{\mathrm{MES}})$ with regard to the sample size to determine how many samples are required for acceptable error margins. The position-space distribution can be obtained directly in discretized form with $L^2$ different outcomes, whereas the momentum-space distribution is continuous in $k_1$ and $k_2$ and needs to be processed via Monte Carlo integration, demanding more samples. We therefore put special emphasis on the momentum distribution in the following and fix the number of position-space samples to $N_\mathrm{pos} =\num{1e4}$.

\begin{figure}
    \centering
    \includegraphics[width=\columnwidth]{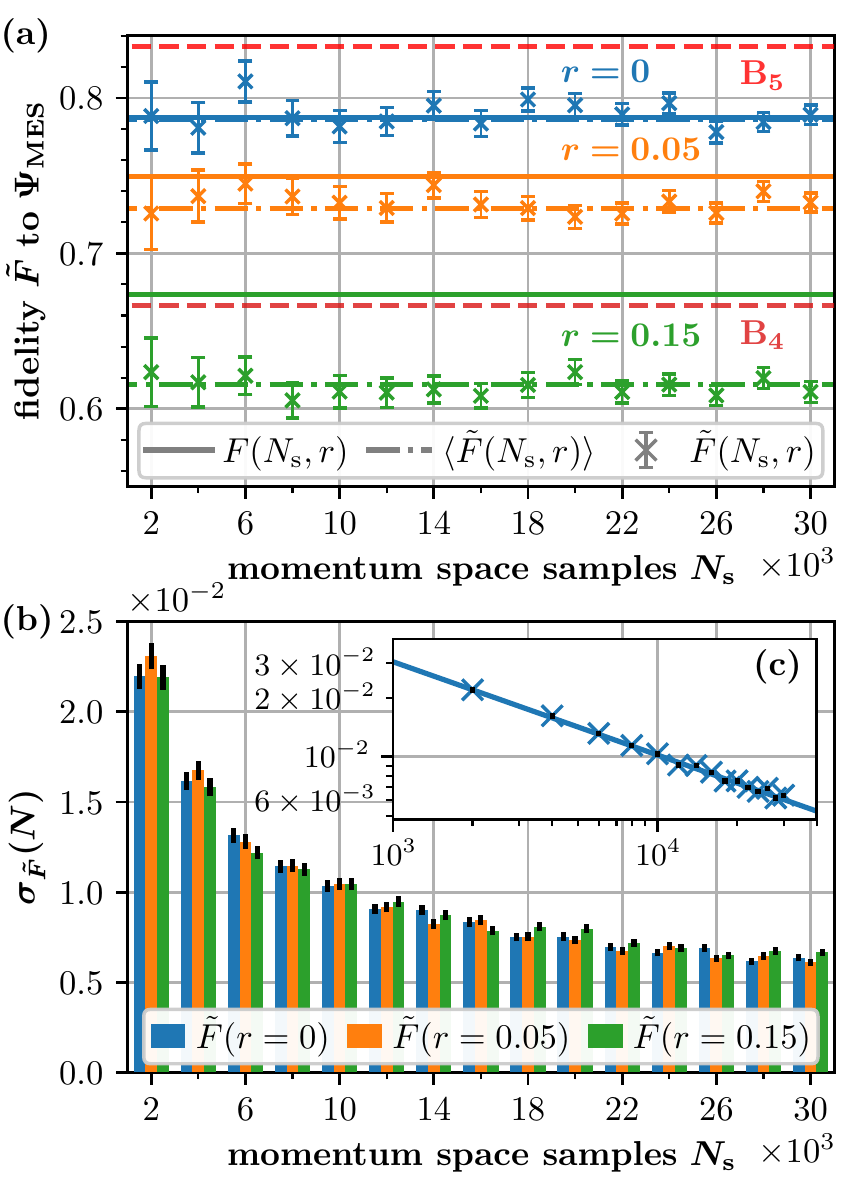}
    \caption{\textbf{(a)} The fidelity bound $\tilde{F}$ as a function of the number of momentum-space samples for mixing strengths \hbox{$r\in\{0,~0.05,~0.15\}$}. The gap between the true fidelity $F$ (solid lines) and the fidelity-bound average $\langle \tilde{F}\rangle$ (dash-dotted lines) increases with growing impurity. Each marker represents one sampling realization. \textbf{(b)} The dependence of the SE of the fidelity bound $\sigma_{\tilde{F}}$ on the number of momentum-space samples $N_s$ for mixing strengths from (a). \textbf{(c)} The linear regression of log-log-represented $\sigma_{\tilde{F}}$ data for $r = 0$ from (b), demonstrating power-law scaling.}
    \label{fig:statistics}
\end{figure}

To analyze scaling properties with regard to available measurement statistics, we compute the fidelity bound $\tilde{F}$ for a wide range of synthetic momentum-space sample sizes $N_\mathrm{s}$. The results for different values of $r\in\{0,0.05,0.15\}$ are presented in Fig.~\ref{fig:statistics}(a). For $\bar{p}=r=0$ (blue data set), the average of the distribution (dash-dotted line) and the true state fidelity coincide, indicating that we can reconstruct the right fidelities without bias. The SE $\sigma_{\tilde{F}}$ of the distribution for different impurities and sample numbers is shown in Fig.~\ref{fig:statistics}(b). We report no significant dependence of $\sigma_{\tilde{F}}$ on the impurity and find a power-law behavior with exponent $b = {(-0.48\pm0.02)}$ [Fig.~\ref{fig:statistics}(c), computed with the $r=0$ data set], consistent with the expectation of Monte Carlo error scaling $\sigma_{\mathrm{MC}}\sim 1/\sqrt{N_\mathrm{s}}$. The complete set of all fitting parameters for this and all following numerical fits can be found in Appendix~\ref{app:fit-parameters}. At high momentum-space sample numbers, we observe a saturation of the error, as the number of position-space samples has been kept constant and the corresponding statistical fluctuations start to dominate. We conclude that \num{1e4} position-space samples and \num{1.2e4} momentum-space samples are sufficient to reduce the SE to $\sigma_{\tilde{F}} < 0.01$, independent of the state impurity.

 When the fidelity lower bound is used to certify the entanglement dimension of the experimental state, the statistical requirements for faithful certification solely depend on the distance to the next threshold value B$_k$. Fidelity-bound values directly in the middle of two B$_k$ lines maximize this distance and have the highest admissible margin of error, whereas fidelities close to thresholds call for ever-increasing sample sets to provide the needed accuracy. The measured bound value can be monitored on the fly to adapt the number of samples taken in order to fulfill the statistical requirements. 
 
 Our data indicate that surprisingly low sample numbers can be sufficient for robust entanglement detection. For example, the distance to the next relevant threshold B$_5$ (red dashed line in Fig.~\ref{fig:statistics}(a) for the pure ground state at $U/J = -12$ is given by $\mathrm{B}_5 - \langle\tilde{F}\rangle > 2 \sigma_{\tilde{F}}$, even at only $N_\mathrm{s} = 2000$ samples, a statistically significant statement. The fidelity to the MES drops with decreasing attractive interaction strength and, with it, the distance to the next lower fidelity threshold. The sample set sizes should thus be adapted for less attractive interaction strengths.

\subsection{Dephasing noise}
\label{sec:dephasing}

\begin{figure}
    \centering
    \includegraphics[width=\columnwidth]{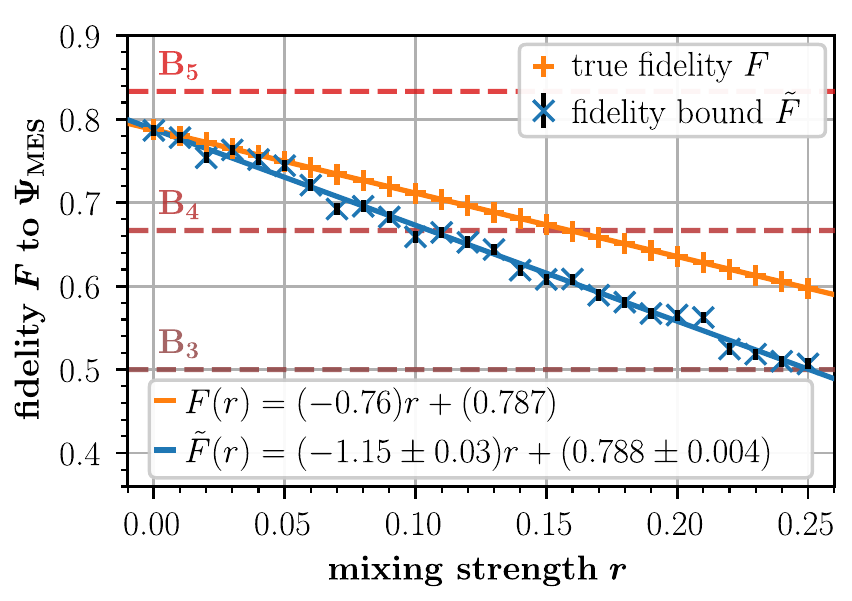}
    \caption{The systematic linear scaling of both $F$ and $\tilde{F}$ with the state-mixing parameter $r$. Since the true fidelity $F$ is computed exactly, linear regression errors $[\mathcal{O}(\num{1e-16})]$ are solely caused by machine precision and are omitted here. The error of $\tilde{F}$ is of a statistical nature and is barely visible.}
    \label{fig:dephasing}
\end{figure}

 Both the fidelity $F$ and the fidelity bound $\tilde{F}$ decrease linearly with the mixing parameter $r$, as shown in Fig.~\ref{fig:dephasing}. The bound declines faster than the actual state fidelity; the linear fit slopes are \hbox{$a = -0.76$} (fidelity) in contrast to \hbox{$\tilde{a} = -1.15\pm0.03$} (fidelity bound) \footnote{The state fidelity is not subjected to any random noise, so the fit errors are at floating point precision and can be neglected.}. Consequently, the tightness gap widens linearly with a slope of $a_{\mathrm{Gap}} = 0.39\pm0.03$ with $r$. Our bound certifies the same entanglement dimension as could be certified with the actual fidelity for most of the investigated impurity regime $r\leq0.25$ and with only one dimension less in the regime $0.1\lesssim r\lesssim0.16$ (Fig.~\ref{fig:dephasing}). Certification of high-dimensional entanglement therefore remains possible even for significantly mixed states. One concrete decoherence effect potentially arising during state preparation is the presence of a thermal bath, resulting in a Gibbs thermal state with finite temperature, i.e., a mixture of ground and excited states. We discuss this case in Appendix~\ref{app:thermal}, finding that our certification scheme is quite robust in the sense that the fidelity bound remains tight up to rather high temperatures. Thus, the generic white noise considered here may well overestimate the typical impact of decoherence on the proposed method.

\subsection{Lattice-potential disorder}
\label{sec:lattice}

Next, we investigate the tightness of our fidelity bound in the presence of lattice-potential disorder,
which typically arises in experiments with arrays of optical tweezers where the relative intensities, and thus the depths, of the individual tweezer traps are difficult to stabilize. We introduce a modified Hamiltonian based on Eq.~\eqref{eq:Hamiltonian} including a normally distributed potential depth fluctuation for each lattice site,
\begin{equation}
    \hat{H}_{\Delta_V} = \hat{H} + \sum\limits_{i}\Delta V_i(\hat{\mathrm{n}}_{i\downarrow}+\hat{\mathrm{n}}_{i\uparrow})\,,\quad \Delta V_i\sim\mathcal{N}\left(0,({J\sigma_V})^2\right),\label{eq:Hfluc}
\end{equation}
with the tunneling strength $J$ as the energy scale. It should be noted that the fluctuations are modeled to be uncorrelated, a realistic assumption in the case of optical tweezer arrays but not necessarily for optical lattices.
Imperfections in the potential landscape cause localization in the ground-state wave function and decreased fidelity to the reference state $\Psi_{\mathrm{MES}}$, as seen in Fig.~\ref{fig:lattice-config}(c). The resulting composition of the localized state is quite different compared to that of $\Psi_{\mathrm{MES}}$, with strongly peaked double-occupation probabilities around some localization center. 
Even though the two distributions differ significantly, our method still enables one to certify an entanglement dimension of $D_{\mathrm{ent}}=4$, demonstrating the wide applicability of the protocol, as we can detect major components of the entanglement spectrum of a state not close to the reference.
In particular, it allows us to track the reduction in entanglement due to disorder-induced pair localization, as we discuss in the following.

\begin{figure}
    \centering
    \includegraphics[width=\columnwidth]{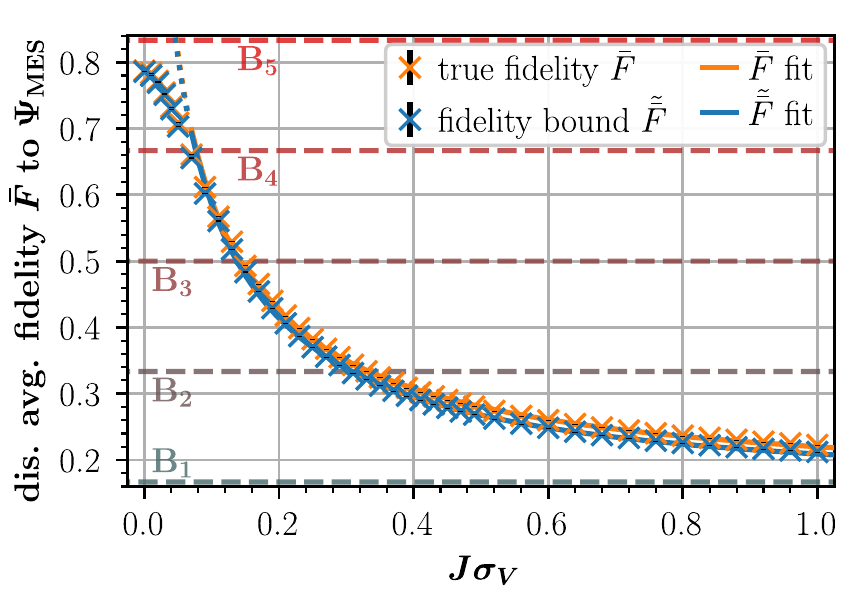}
    \caption{The disorder ensemble average of the fidelity $\bar{F}$ as a function of the strength (standard deviation $J\sigma_V$) of lattice depth fluctuations. After an initial transitional phase, the fidelity and the bound follow a stretched exponential decay for potential fluctuations with $J\sigma_V \gtrsim 0.07$. The data points in the dotted part of the line are excluded from the fit, indicating deviating behavior for very weak fluctuations due to finite-size effects.}
    \label{fig:lattice_disorder}
\end{figure}

In the strongly attractive regime of $U/J = -12$, the states with both atoms at the same lattice site make a contribution of $95.7\%$ to the pure undisturbed ground-state populations. It is therefore a reasonable simplification to treat the atom pair as a bound dimer moving through the lattice. The ground-state localization is then in agreement with the predictions of Anderson localization for disordered potentials, where the occupation probability is suppressed exponentially when going away from the localization center \cite{Anderson1958}. 1D systems are expected to localize for any nonzero potential disorder, with the localization length depending on the disorder strength \cite{Abrahams1979}. 

\begin{figure*}
    \centering
    \includegraphics[width=\textwidth]{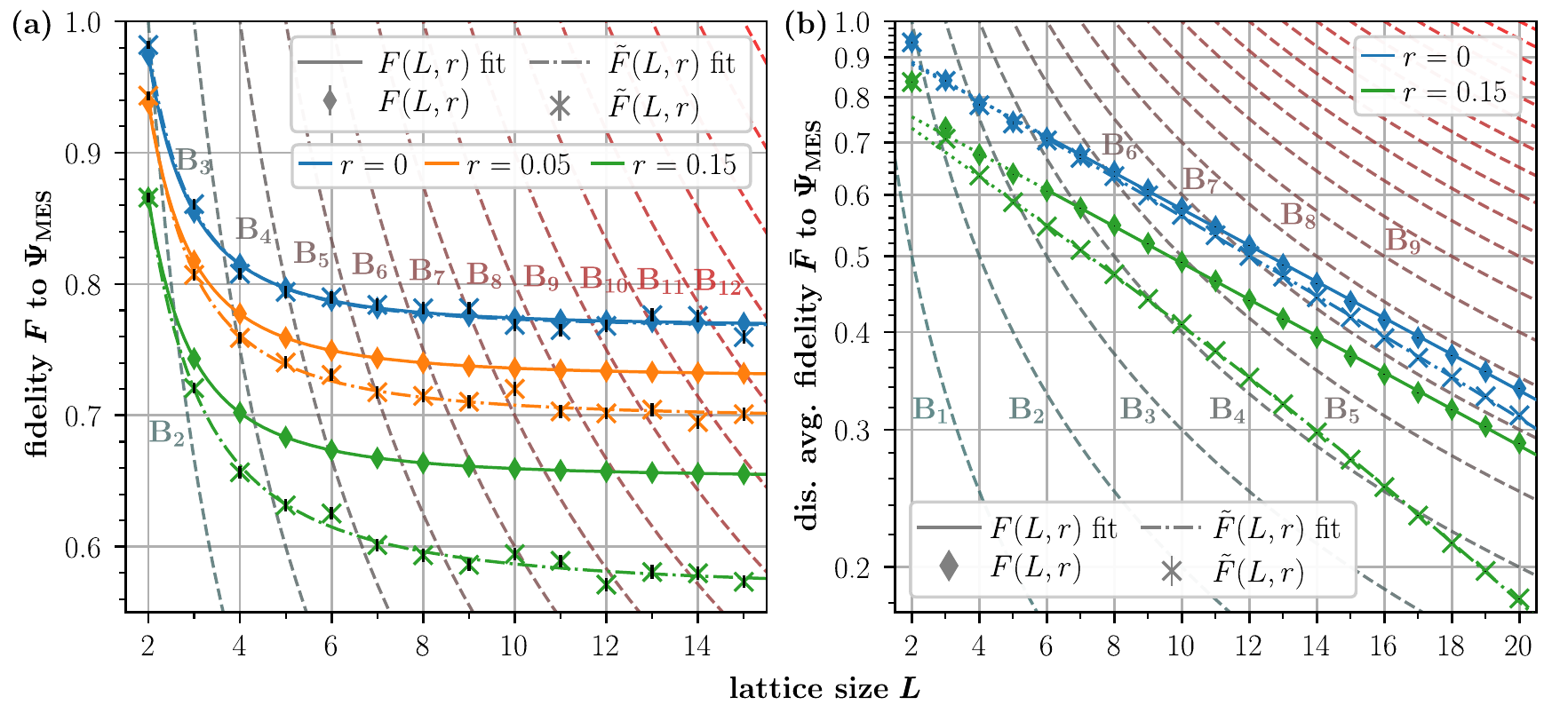}
    \caption{The lattice-size dependence of the fidelity. \textbf{(a)} The scaling of the fidelity $F$ of the ground state of a flat optical lattice as a function of the number of lattice sites $L$ and the mixing parameter $r$. The numerical data, including statistical noise, fit well to the asymptotic algebraic behavior. The certifiable entanglement dimension continues to grow with increasing lattice size, as the fidelity asymptotically approaches its infinite-system-size value. \textbf{(b)} A log-linear plot of the fidelity for a disordered lattice of size $L$ with fixed disorder strength $J\sigma_V = 0.05$, showing an exponential fidelity decay. The number of entanglement dimensions accessible to certification has a maximum of $D_{ \mathrm{ent}} = 7$ before decreasing again with growing system size. The numerical fits have been computed using data points with $L\geq6$ (nondotted lines) and contain an offset for the bounds. The data for $r=0.05$ are not included in the figure for better visual clarity. The statistical error bars, especially on the true fidelity $F(L,r)$, are barely visible.
   Both configurations are evaluated with an increased \num{2.5e4} position-space and \num{5e4} momentum-space samples.}
    \label{fig:L-both-dependence}
\end{figure*}

To investigate the effect of shot-to-shot lattice-potential fluctuations on the (detected) MES fidelity, we simulate single experiments on ground-state mixtures of \num{1e3} individual disorder realizations configured according to Eq.~\eqref{eq:Hfluc} and compute the disorder ensemble average $\bar{F}$ over \num{1e3} experimental runs. Both $\bar{F}$ and $\tilde{\bar{F}}$ decrease according to a stretched exponential law  $\propto \exp{-b(J\sigma_V)^c}$ with increasing potential depth fluctuation $J\sigma_V$ and approach the $\mathrm{B}_1$ entanglement threshold, shown in Fig.~\ref{fig:lattice_disorder}.
The bound tightness does not decrease significantly compared to the disorder-free lattice, even for the strongest simulated fluctuation strengths. This is quite remarkable, as each investigated state is a mixture of thousands of individually localized disorder realizations. Consequently, our bound certifies the same entanglement dimension or Schmidt number as the true fidelity for a large regime of disorder strengths. This behavior is markedly different from dephasing noise, where we have found a linearly widening gap between fidelity and bound.

The breakdown of the fit at small disorder strengths is caused by the finite size of the lattice. For very weak disorder, the localization length exceeds the lattice scale. In this regime, the fidelity therefore only decreases slowly with increasing disorder strengths, up until single disorder centers can be fully resolved in the lattice. The $J\sigma_v=0$ data point additionally marks the critical point of the localization phase transition in 1D, so anomalous behavior is expected here. Consequently, small disorder strengths do not significantly decrease the fidelity $F(\hat{\rho}, \Psi_{\mathrm{MES}})$, and thus our bound also decreases at a reduced rate.

\subsection{Lattice-size dependence of the state fidelity}
\label{sec:scalability}

The scalability of entanglement certification with respect to the lattice size $L$ is of significant concern for experimental implementations. To systematically investigate this, we repeat our previous simulation for a range of different lattice sizes. Our data for a system with finite sampling statistics and dephasing noise show an algebraic asymptotic decline of the fidelity with growing lattice size, as shown in Fig.~\ref{fig:L-both-dependence}(a) (the fit model and parameters are given in Appendix~\ref{app:fit-parameters}). The fidelities and fidelity bounds approach constant nonzero values for $L\rightarrow\infty$, depending on the state-mixing strength $r$. Consequently, certified entanglement dimensions continue to grow as $L\rightarrow\infty$.
We find that the scaling behavior of our bound depends on the level of dephasing noise.

The addition of lattice disorder changes the situation. From previous data (see Sec.~\ref{sec:lattice}), we expect localization into dimers but the dependence on the lattice size is not immediately evident. Our investigation of a lattice with fixed disorder strength $J\sigma_v=0.05$ yields an exponential fidelity drop-off $\sim\exp(-bL)$, as shown in Fig.~\ref{fig:L-both-dependence}(b). As our investigated states are mixtures of ground states of different disorder realizations, we do not expect the bound to be tight in the limit of large $L$. To account for this, we include an offset $c$ in the exponential fit to $\tilde{F}$.
All fits match the data very well at large lattice sizes but significantly underestimate the fidelity in double- and triple-well configurations. Again, finite-size effects offer a plausible explanation for this behavior: the localization length of the system can exceed the lattice size, making it impossible to resolve a localization center fully in small systems.

Since the entanglement-dimension thresholds scale only linearly, B$_k\sim~L^{-1}$, as compared to the exponentially decaying fidelity, the certified entanglement dimension decreases to $D_{\mathrm{ent}} = 1$ for $L\rightarrow\infty$.  Consequently, after an initial increase of certifiable entanglement, the entanglement dimension accessible through the bound starts to decline. Based on the reported fit, we extrapolate a maximum certifiable entanglement dimension of $D_{\mathrm{ent}} = 7$ for pure states and $D_{\mathrm{ent}} = 5$ for $r=0.15$ for a fixed disorder standard deviation of $J\sigma_v = 0.05$. 

The above simulations demonstrate the strong impact of site-to-site lattice-potential fluctuations on the scaling behavior of fidelity and thus on the certifiable entanglement dimension for large lattice sizes. While in the case of a disorder-free lattice the certifiable entanglement dimension increases indefinitely with lattice size, disorder-induced localization effects lead to a maximal certifiable dimension reached at some finite lattice size, depending on the disorder strength. We note that site-to-site potential fluctuations may be present for arrays of optical tweezers, while in the case of optical lattices, realized by a single retroreflected laser beam, intensity fluctuations will lead to correlated potential fluctuations not affecting the ground-state properties. Also, the precise properties of the prepared state may depend on the experimental preparation scheme, not discussed in this work. Furthermore, viewing disorder as a feature and tuning its strength deliberately, our method allows the study of dimer localization through the lens of the entanglement spectrum.

\begin{figure}
\centering
    \includegraphics[width=\columnwidth]{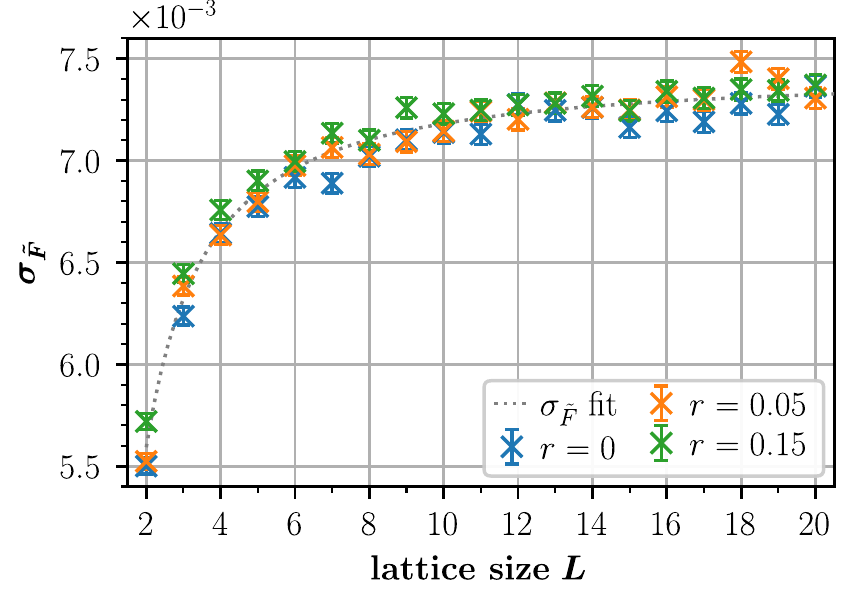}
    \caption{The dependence of the SE of the fidelity bound $\sigma_{\tilde{F}}$ on the lattice size $L$ at a fixed number of samples. All mixing rates $r\in\{0,0.05,0.15\}$ show similar initial growth of $\sigma_{\tilde{F}}$ before saturation. The combined data set is well described using an algebraic asymptotic growth model, showing little variation in $\sigma_{\tilde{F}}$ for lattices with $L\gtrsim10$ }
    \label{fig:L-resampling}
\end{figure}

Finally, we consider the dependence of the statistical errors on the lattice size, again with fixed sample numbers. For this purpose, we analyze $\num{1e4}$ bootstrap resampling realizations to estimate the SE $\sigma_{\tilde{F}}$ for lattice chains with lengths up to $L=20$, as displayed in Fig.~\ref{fig:L-resampling}. All three simulated mixing rates give qualitatively and quantitatively similar errors. To extract the general trend, we average over the three mixing-rate data sets for improved statistics and fit with an algebraic asymptotic growth model, which shows good agreement in the investigated regime (the model is also listed in Appendix~\ref{app:fit-parameters}). Therefore, we find $\sigma_{\tilde{F}}$ to be largely independent of the lattice size. Scaling up to extended lattice chains is therefore not statistically prohibitive, opening up the possibility of feasibly preparing and certifying states with very high-dimensional entanglement.

In summary, state dephasing and lattice fluctuations have different impact signatures on both the true fidelity $F$ and on our fidelity bound $\tilde{F}$. While the bound tightness is loosened by growing dephasing, with a linearly widening gap between $F$ and $\tilde{F}$, it remains mostly tight in the presence of lattice-potential fluctuations. The statistical errors follow the expected Monte Carlo scaling $\propto 1/\sqrt{N_{\mathrm{s}}}$; very moderate sample numbers of $\approx \num{1e4}$ both in momentum and position space are sufficient to reduce the SEs to the subpercent range for all investigated lattice sizes. The bound is therefore robust with respect to typical noise sources and the entanglement-certification capability comes close to that of the true fidelity.

\section{Multiple Particles per Species}
\label{sec:ML}

In the context of quantum simulation of condensed matter physics problems, the two-atom configuration discussed so far presents a somewhat unphysical low-density limit. 
Eventually, one would like to access the entanglement structure near half-filling, meaning atom number $N=L/2$, where true many-body effects emerge.

However, our method still relies on the measurement of coefficients of trigonometric basis functions, the number of which scales with the local Hilbert space size and thus exponentially in the particle number. Hence the true many-body regime stays out of reach for the scheme presented in this work.
Nonetheless, studies of few-body cold-atom systems in the past decade have revealed that the few-body dynamics approache the many-body limit even at very moderate particle numbers \cite{Wenz2013, Rammelmuller2017}. Few-body systems are thus interesting candidates for quantum simulation and, in particular, entanglement certification, and give experimentalists capabilities beyond that of simpler two-particle systems such as entangled photon pairs. Recent success in the preparation and control of indistinguishable atom systems motivate this ansatz \cite{Becher2020, Klemt2021}. Here, we want to extend our method to multiple particles in each of the two subsystems and present numerical simulations for up to $N=4$ particles per species.

\subsection{Theoretical considerations}
\label{sec:MLtheory}

Systems in which the number of atoms per species is increased to $N>1$ can conveniently be described in a second quantization picture with different Fock modes. These modes are labeled by the spin of the particles and their lattice positions and are occupied by a given number of particles.
For the fermionic atoms in the Fermi-Hubbard model, each lattice site can at most be populated by one atom per species due to Pauli exclusion. Hard-core bosons have the same exclusion rule, here enforced by strong repulsive on-site interactions. The resulting dimension of the local Hilbert space, i.e.,\ the Hilbert space available to each species, which determines the maximal entanglement dimension, thus becomes
\begin{equation}
    D^{\mathrm{max}}_{\mathrm{ent}} = \binom{L}{N}\,.\label{eq:entdimML}
\end{equation}
A half-filling configuration gives the highest-possible entanglement dimension for a given lattice size, with \hbox{$D^{\mathrm{max}}_{\mathrm{ent}}(N=L/2) = L!/[(L/2)!]^2$}. The $2N$-body wave function $\ket{\Psi}^{N+N}_{\mathrm{MES}}$ then reads
\begin{equation}
\ket{\Psi}_{\mathrm{MES}}^{N+N} = \frac{1}{\sqrt{\binom{L}{N}}}\!\sum_{{\substack{m_i=1\\m_i<m_{i+1}}}}^{L}\!\ket{m_1\dots m_N}_{\mathrm{A}} \otimes \ket{m_1\dots m_N}_{\mathrm{B}},\label{eq:MES_ML}
\end{equation}
with the normalization adapted to reflect the changed Hilbert-space size. In this notation $\ket{m_1\dots m_N}_{\mathrm{A/B}}$ designates the Fock state of species A or B, where sites $m_1\dots m_N$ are occupied by one atom each. Here, the $m_i$ are in ascending order and are mutually different due to the aforementioned exclusion rules.

The general approach of bounding the fidelity to the MES to bound the Schmidt number remains unaltered. All experimental tools for single-atom and spin-resolved detection are still applicable but one has to take care to properly address the indistinguishability within the subspecies.  While state populations can be extracted in a straightforward extension to the two-atom case, some changes have to be applied to access the coherences in Eq.~\eqref{eq:fidelsplit} in the second quantization picture. Here, due to different commutation relations of fermions and bosons, our bound behaves differently for the two cases. In this work, we focus only on hard-core bosons and fermions, as their Hilbert spaces are identical and can thus be treated analogously. The detailed construction of the fidelity lower bound from multiparticle real-space and momentum correlation functions is presented in Appendix~\ref{app:indis}. The crucial difference between fermions and bosons is the appearance of signs in the coherence terms stemming from the fermionic anticommutation relations. This diminishes the tightness of the estimate used in Eq.~\eqref{eq:fidelbound}, even for pure states, and makes high-dimensional entanglement certification more challenging for fermionic systems than for hard-core bosons, as we show in Sec.~\ref{sec:MLresults}.

Lastly, we briefly address the scalability of the method toward larger particle numbers. Increasing the system size in terms of the number of atoms in the system requires significant computational resources, both for synthetic data generation and data processing. Furthermore, the necessary measurement statistics also increase for systems with higher atom counts. Our data processing is based on Monte Carlo techniques, which do not inherently scale with the dimension of the momentum space, i.e.,\ the number of atoms in the system, but scaling can be introduced through the variance of the joint momentum distribution. We investigate these statistical scaling properties in \hbox{Sec.~\ref{sec:finitesize}}.

\subsection{Numerical results}
\label{sec:MLresults}

\begin{figure}
    \centering
    \includegraphics[width=\columnwidth]{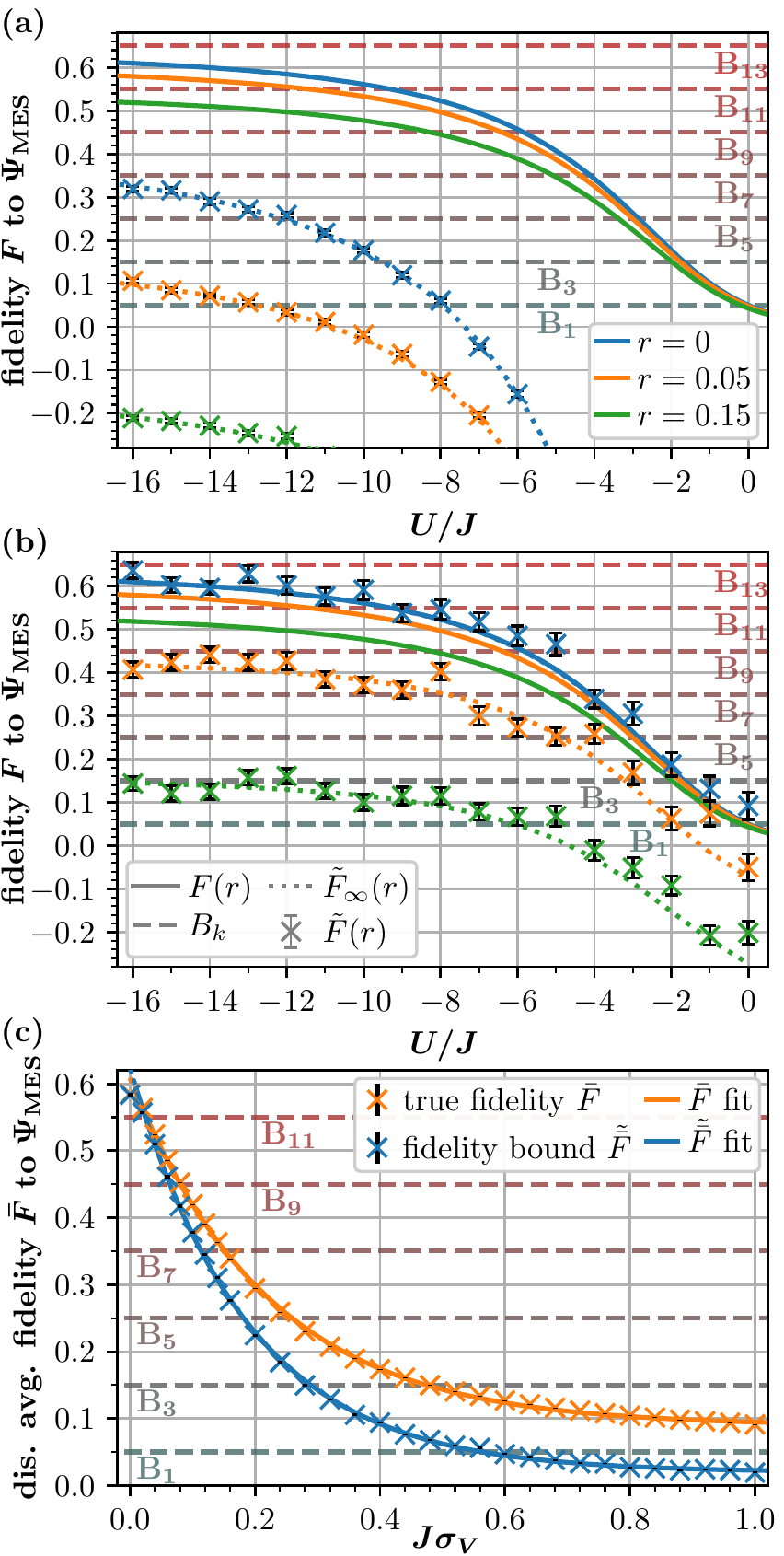}
    \caption{The numerical results for the entanglement-dimension certification of $3+3$ indistinguishable atoms in a lattice with $L=6$. Scaling of the fidelity $F$ and the fidelity bound $\tilde{F}$ for different interaction-to-tunneling-strength ratios $U/J$ for pure ($r = 0$) and dephased ($r \in \{0.05,~0.15\}$) states with \textbf{(a)} fermions and \textbf{(b)} hard-core bosons. The dotted line represents the infinite-measurement-statistics limit $\tilde{F}_{\infty}$ computed using exact coherences of $\hat{\rho}$. \textbf{(c)} The scaling of the disorder ensemble average $\bar{F}$ as a function of the normalized optical-lattice depth fluctuation $J\sigma_v$ for the pure ground state of hard-core bosons. We find good agreement with exponential decay for both the true fidelity $\bar{F}$ and our bound $\tilde{\bar{F}}$ for $J\sigma_v\geq0.01$ disorder strengths. The simulation has been conducted at $U/J=-12$. All measurements have been simulated using \num{1e5} momentum-space and position-space samples each.}  
    \label{fig:fidelityML}
\end{figure}

We simulate the ground state of $N=3$ particles of both species in a lattice with $L=6$ for both fermions and hard-core bosons. This setup enables a maximum entanglement dimension of  $D^{\mathrm{max}}_{\mathrm{ent}} =\binom{6}{3}=20$ [see Eq.~\eqref{eq:entdimML}]. To compare the behavior of this few-body system with that of two atoms, we repeat the interaction-strength sweep shown in Fig.~\ref{fig:fidelity}. Our numerical data show that the fidelity to the MES in the strongly attractive regime is lower than in the two-atom case [see Figs.~\ref{fig:fidelityML}(a) and \ref{fig:fidelityML}(b)] but the behavior is otherwise qualitatively the same. The fidelity reduction is caused by a combination of same-site exclusion, which increases the distance between atoms of the same species, and finite-size effects, penalizing occupation of sites close to the edges. The combination of both effects leads to a very nonuniform distribution of dimer populations. However, as anticipated, the fidelity bound $\tilde{F}$ shows a strong dependence on the underlying quantum statistics; for bosons, much higher and thus tighter fidelity bounds were achieved compared to fermions. This also leads to a large difference in terms of the certified entanglement dimension; in the pure case at $U/J=-15$, we certify $D_{\mathrm{ent}}=7$ for fermions and $D_{\mathrm{ent}}=13$ for hard-core bosons. The trend also carries over to dephased states with $r>0$, where we observe a stronger impact of dephasing than in the two-atom case. In the fermionic case, at the strongest investigated dephasing of $r=0.15$, no entanglement is witnessed in the ground state and only $D_{\mathrm{ent}}=2$ is found for weaker dephasing of $r=0.05$. The impact is less severe for bosons, where we find $D_{\mathrm{ent}}=3$ for $r=0.15$ and $D_{\mathrm{ent}}=13$ for $r=0.05$, respectively. These findings contrast with our data for a half-filling configuration for $N=2$ atoms per species with $L=4$, shown in Fig.~\ref{fig:2+2} in Appendix~\ref{app:2+2}. Here, bosons and fermions show very similar results, with minor deviations only visible for pure states in the weakly attractive regime. Additionally, the effect of dephasing is much more comparable to our initial findings for $N=1$ atom per species in Fig.~\ref{fig:fidelity}. With higher numbers of particles and lattice sites present in the system, an increasing amount of coherences have to be subtracted using the bound in Eq.~\ref{eq:fidelbound}, explaining the difference in performance for different system sizes.

Finally, we also investigate the case of $N=4$ hard-core bosons per species on a lattice with $L=8$. Using \num{3e5} samples in both position and momentum space, we are able to estimate $\tilde{F}=0.50\pm0.06$ at $U/J = -15$. Given $D^{\mathrm{max}}_{\mathrm{ent}} = \binom{8}{4}=70$, this fidelity translates into a certified entanglement dimension with respect to a $1\sigma$ confidence interval of  $D_{\mathrm{ent}}=31$ ($3\sigma$ confidence: $D_{\mathrm{ent}}=23$). In accordance withe the above-described dephasing characteristics at $r=0.05$, we see a strongly reduced fidelity bound of $\tilde{F}=0.10\pm0.07$, which gives $D_{\mathrm{ent}}=3$ at the $1\sigma$ level.

Interestingly, the addition of disorder on the lattice reveals some key differences compared to the $1+1$-atom case. Instead of the stretched exponential approach toward $\mathrm{B}_1$ we find standard exponential decay of $\bar{F}$ with the bound decreasing significantly below the entanglement-detection threshold [cf.\ Fig.~\ref{fig:fidelityML}(c)]. The matter wave function cannot converge to one localization center but is distributed across the entire lattice due to the above-discussed exclusion rules. A nonzero number of states with unpaired atoms retain nonvanishing populations and connected coherences, which in turn induce tightness loss. Nonetheless, the initial transitional phase is comparatively short, with a good numerical fit agreement already for $J\sigma_{V}\geq0.01$. To summarize, the onset of localization effects is found for weaker disorder in few-body systems and quickly converges toward the sensible infinite-disorder ensemble, i.e., the perfect mixture of localized dimer states.
However, state dephasing is the dominant effect, as even the strongest disorder strength $J\sigma_v=0.15$ results in a deviation $\Delta_F \coloneqq |F-\tilde{F}| \approx 0.08$, half of the difference caused by minor dephasing at $r=0.05$.

\subsection{Scaling of sampling complexity with \texorpdfstring{$N$}{N}}
\label{sec:finitesize}

\begin{figure}
    \centering
    \includegraphics[width=\columnwidth]{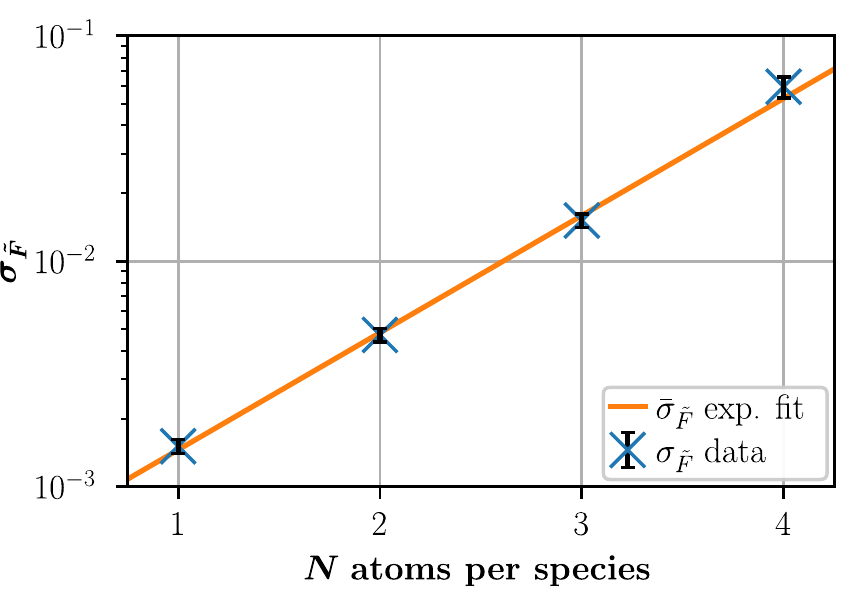}
    \caption{A Log-linear plot of the fidelity bound SE $\sigma_{\tilde{F}}$ for \hbox{$N+N$} atoms in a lattice at half-filling ($N=L/2$). For coherence reconstruction, \num{3e5} position- and momentum-space samples each were used. The SE is well described using an exponential numerical fit. Because of the higher computational cost for $4+4$ atoms, a smaller number of resamples was taken, leading to higher uncertainty in $\sigma_{\tilde{F}}$.}
    \label{fig:finitesize}
\end{figure}

To gain a better understanding of the complexity in terms of the required experimental runs $N_{\mathrm{s}}$, we conducted a scaling analysis of $N+N$ atoms at half-filling for $N\in\{1,2,3,4\}$. We have found that the SE of the fidelity bound $\sigma_{\tilde{F}}$ increases exponentially with $N$ (see Fig.~\ref{fig:finitesize}). From the fit coefficients one can extract an expected increase of $\sigma_{\tilde{F}}$ by a factor $s_{N} = 3.31\pm0.13$ for every additional atom pair introduced into the system. Using the earlier-confirmed (cf.\ Fig~\ref{fig:statistics}) $\sigma_{\tilde{F}}(N_{\mathrm{s}})\sim1/\sqrt{N_{\mathrm{s}}}$ relation, an increase of the sample size by a factor of $s_{N}^2 = 11.0\pm0.9$ is necessary to keep statistical errors constant while increasing the particle number per species, $N$, by one.
It shows that scaling deep into the many-body regime remains infeasible but configurations with a few atoms per species are realistically achievable.

\section{Multipartite Entanglement}
\label{sec:multipartite}

In the previous section, we showed that when extended to few-body systems of multiple atoms per atomic species, our method still succeeds in the certification of high-dimensional entanglement, even for mixed states. Since only two spin states are populated, the system is fully described by the atom number $N$, the interspecies interaction strength $U$, and the tunneling strength $J$, so the same experimental toolbox can be used as for the case of one atom per species.
When, instead, a higher number of spin states and thus entanglement parties is involved, a plethora of experimental and theoretical complications arise for entanglement detection but we can adapt the bound $\tilde{F}$ to be able to certify genuine multipartite entanglement. In the following, we first formulate the theoretical framework needed for the classification of multipartite entanglement in this system and then describe a possible setup certifying high-dimensional tripartite entanglement. Finally, we present simulation results of entanglement certification for three atomic species in an optical lattice of $L=6$ sites.

\subsection{Multipartite-entanglement certification}
While bipartite entanglement of pure states is fully developed theoretically, many questions are still open concerning the characterization and certification of multipartite entanglement. For states consisting of three entangled qubits, two sets of nonequivalent states sharing genuine tripartite entanglement have been identified, those equivalent under local operations and classical communication (LOCC) to the Greenberger-Horne-Zeilinger (GHZ) state and those LOCC equivalent to the $W$ state \cite{Greenberger1989, Dur2000, Horodecki2009}.  Because of this nonequivalence of entanglement, the Schmidt decomposition is no longer defined for general multipartite states. Different methods are therefore needed to obtain and describe the entanglement structure of multipartite quantum states.  Numerous different approaches have been proposed to define canonical forms of tripartite and multipartite states with a minimal number of nonzero coefficients. However, to uniquely define any given quantum state through these methods, a number of parameters significantly higher than the local Hilbert-space dimension is required  \cite{Acin2000, Carteret2000, Huber2013a, Huber2013b}. For some states, most notably also for generalizations of the GHZ state to higher local dimensions,
\begin{equation}
    \ket{\psi}_{\mathrm{ABC}} = \sum\limits_{\mathclap{i=1}}^{k}\lambda_i \ket{i}_{\mathrm{A}}\otimes\ket{i}_{\mathrm{B}}\otimes\ket{i}_{\mathrm{C}}\label{eq:tripartite},
\end{equation}
it is still possible to define a generalized Schmidt decomposition, as every contribution to $\ket{\psi}_{\mathrm{ABC}}$ combines orthogonal basis vectors $\ket{i}$ for all three subsystems, assumed to have the same local Hilbert-space dimensions. No basis transformation can therefore reduce the number of terms used for the representation of Eq.~\eqref{eq:tripartite} any further \cite{Horodecki2009, Thapliyal1999}. One can now define a multipartite-entanglement dimension with the properties of an entanglement monotone in analogy to that of bipartite states \cite{Chen2017}, also with a maximum value of \hbox{$D^{\mathrm{max}}_{\mathrm{ent}} = \mathrm{dim}\,\mathcal{H}_{\mathrm{A}}$}. A generalized GHZ state with equal contributions on all sites given by
\begin{equation}
    \ket{\mathrm{GHZ}}_{L} = \frac{1}{\sqrt{L}}\sum_{{m=1}}^{L}\ket{mmm}\label{eq:MES_tri}
\end{equation}
therefore represents a suitable generalization of the two-atom MES [Eq.~\eqref{eq:MES}] as a reference state. It should be noted that this state is not maximally entangled in the sense that it has the maximum number of terms needed to be faithfully represented among all states with genuine tripartite entanglement but, rather, has the highest number of terms possible for it to also have a generalized Schmidt decomposition of the given form.

The entanglement dimension of an experimental state $\hat{\rho}$ can be bounded by a set of fidelity thresholds B$_k$ to that reference state analogous to Eq.~\eqref{eq:Bk}, opening up in principle the same certification route taken for bipartite entanglement. We prove these bounds in Appendix~\ref{app:bounds}. The algorithm given by Eqs.~\eqref{eq:pos_space_rep}-\eqref{eq:fidelbound} can be adapted straightforwardly to include three or more atomic species (for details, see Appendix~\ref{app:multipartite}).

\subsection{Experimental model}

\begin{figure}
    \centering
    \includegraphics[width=\columnwidth]{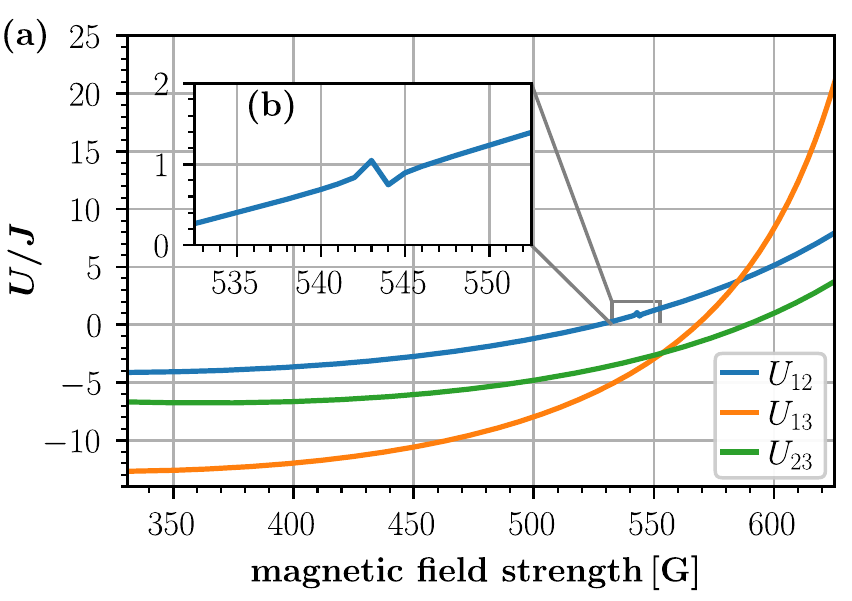}
    \caption{\textbf{(a)} The dependence of the three different interaction-to-tunneling-strength ratios $U_{12}/J$, $U_{13}/J$, and $U_{23}/J$ on the external magnetic field (in gauss) based on Ref.~\cite{Zurn2013} and gauged to fit experimental data published in Ref.~\cite{Bergschneider2019}. \textbf{(b)} Magnification of the narrow $s$-wave Feshbach resonance at 543$\,$G \cite{Dieckmann2002, Schunck2005}.}
    \label{fig:U_tri}
\end{figure}

There are several different possible cold-atom implementations in which multipartite entanglement can be realized. Here, we choose a generalized Hubbard model with three distinct atomic species. The interaction strength between different spin states is usually regulated through the use of a magnetic Feshbach resonance \cite{Feshbach1958, Schafer2020}. When a third spin state is added to the system, each of the three possible atom pairs is now governed by their individual interaction strengths $U_{ij}$. To experimentally realize control over a mixture of three different spin states, an isotope with three overlapping Feshbach resonances connecting three low-energy eigenstates can be used. One possible choice is $^6$Li, for which Feshbach resonances for the three lowest energy states at 690\,G, 811\,G, and 834\,G are experimentally accessible and have been realized before \cite{Ottenstein2008, Huckans2009, Azaria2009}. Since all three Feshbach resonances are magnetic, it is no longer possible to control the individual interaction strengths independently, but, rather, all three values $U_{12}$, $U_{23}$, and $U_{13}$ are tuned at the same time through shifts of the external magnetic field. For field strengths up to $527\,$G, all three scattering lengths are negative, delivering a broad regime of attractive interactions between all atom species, as shown in Fig.~\ref{fig:U_tri}. States close to the MES [Eq.~\eqref{eq:MES_tri}] can thus be realized by preparing the Hubbard-model ground state in this regime. A three-particle extension to the Hubbard-model Hamiltonian can be constructed as
\begin{equation}
    \hat{H} = -J\sum\limits_{\mathclap{\sigma,i}}(\hat{c}_{i,\sigma}^\dagger \hat{c}_{i+1,\sigma}^{\phantom{\dagger}} + \mathrm{h.c.}) + \sum\limits_{\mathclap{\substack{\sigma_1,\sigma_2\\\sigma_1<\sigma_2}}}\,\sum\limits_{i}U_{\sigma_1\sigma_2}\hat{\mathrm{n}}_{i,\sigma_1}\hat{\mathrm{n}}_{i,\sigma_2}\label{eq:hamilton_tri}\,,
\end{equation}
with $\sigma, \sigma_1, \sigma_2 \in \{1,2,3\}$ labeling the different hyperfine states \cite{Azaria2009}. 
We base our numerical simulation of tripartite entangled systems on precise scattering lengths for $^6$Li published in Ref.~\cite{Zurn2013}. From these measurements, we derive $U/J$ values for different magnetic field strengths gauged to fit the interaction-strength data for $U_{13}$ reported in Ref.~\cite{Bergschneider2019} [Fig.~\ref{fig:U_tri}(a)] to establish experimental comparability. This provides access to the interaction-strength triplet for a wide field-strength regime and thus enables one to study high-dimensional tripartite entanglement in Hubbard-model ground states. An alternative approach could be based on ultracold fermionic atoms in optical lattices with SU($N$)-symmetric interactions \cite{scazza_observation_2014, zhang_spectroscopic_2014}. They have recently been shown to feature strong effective multibody interactions, making them a promising atomic platform for the preparation of multipartite entanglement in the future \cite{perlin_effective_2019}.

\subsection{Numerical results}

\begin{figure}
    \centering
    \includegraphics[width=\columnwidth]{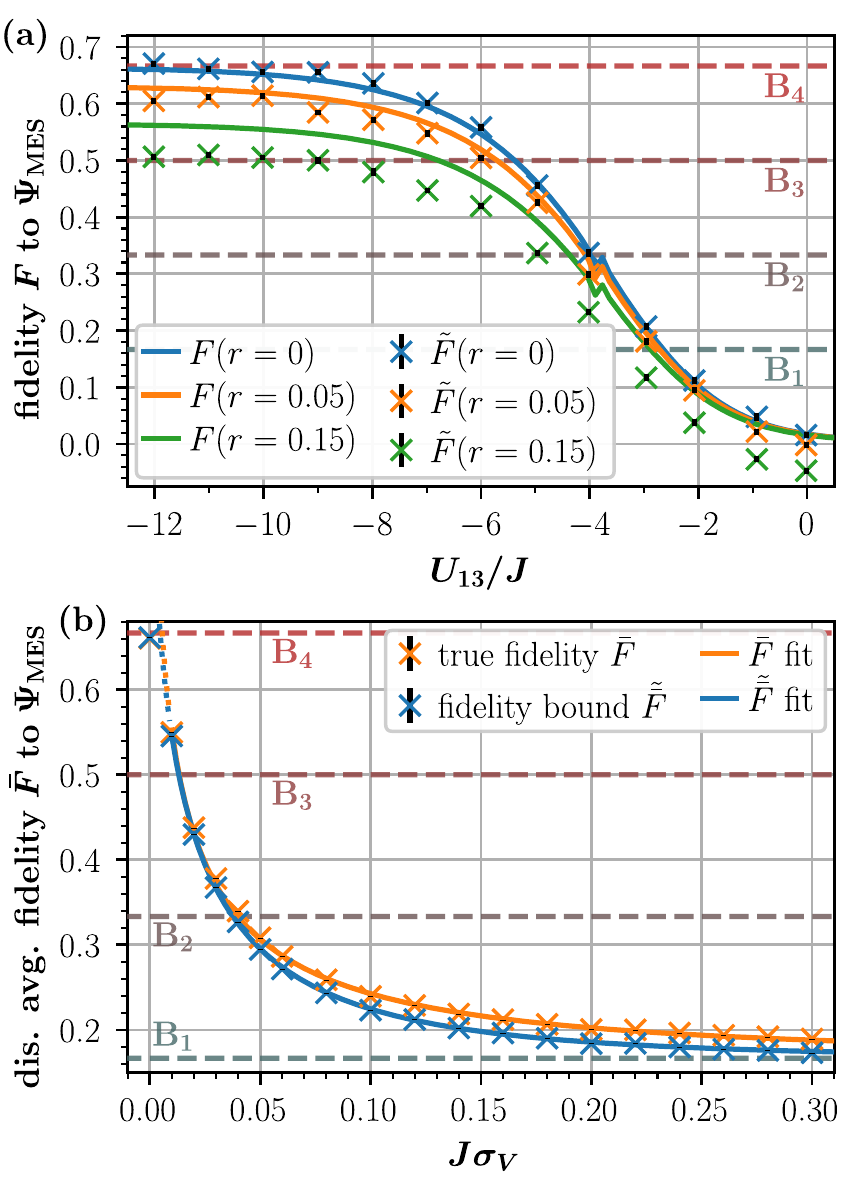}
    \caption{The numerical results for entanglement-dimension certification of a tripartite-state configuration in a lattice with $L=6$. \textbf{(a)} The scaling of the fidelity $F$ and the fidelity bound $\tilde{F}$ for different interaction-to-tunneling-strength ratios $U/J$ for pure ($r = 0$) and dephased ($r \in \{0.05,~0.15\}$) states. \textbf{(b)} The dependence of the disorder-averaged fidelity $\bar{F}$ on the normalized strength of the lattice-potential fluctuations, $J\sigma_V$, for the pure ground state at $U_{13}/J = -12$ ($U_{12}/J = -3.67$, $U_{23}/J = -6.66$). The fidelities are adequately described through a stretched exponential fit approaching the $\mathrm{B}_1$ boundary in the strong-disorder limit. The two simulations used to create the displayed data sets each utilized \num{5e4} momentum-space and \num{1e4} position-space samples.
     The statistical errors caused by disorder averaging and limited statistics are relatively small. Data points in the dotted part of the fit have been excluded from fitting.}
    \label{fig:fidelitytripartite}
\end{figure}

To assess the effect of the new intricate triplet structure of interaction strengths, we perform a sweep across the accessible range of magnetic field strengths $B$ for three distinguishable atoms in the ground state of Eq.~\eqref{eq:hamilton_tri}. The result is presented as a function of $U_{13}$ in Fig.~\ref{fig:fidelitytripartite}(a). All presented true fidelities $F(\hat{\rho},\Psi_{\mathrm{MES}})$ are again computed through exact diagonalization of the Hamiltonian. The signal found at $U_{13}\approx-3.8J$ relates to a narrow $s$-wave Feshbach resonance at $B=523\,$G [magnified in Fig.~\ref{fig:U_tri}(b)], which was earlier reported in Refs.~\cite{Dieckmann2002, Schunck2005}. The observed fidelities are of similar magnitude as values reported for the $3+3$ atom configuration in Figs.~\ref{fig:fidelityML}(a) and \ref{fig:fidelityML}(b) but with significantly higher fidelity bounds for the tripartite configuration. The impact of dephasing is of similar strength, as observed for simple two-atom configurations in Fig.~\ref{fig:fidelity}.

The analysis of lattice disorder reveals differences compared to our results for two-species settings.
At \hbox{$U_{13}/J = -12$}, we see a stretched exponential decay in both $F$ and $\tilde{F}$ with lasting bound tightness, matching our results for two-atom configurations. However, we find a much steeper fidelity reduction and a smaller initial plateau. The fit yields stretch powers of $c = 0.300\pm0.013$ for the true fidelity and $\tilde{c} = 0.489\pm 0.011$ for our fidelity bound. The three strongly attractive interaction strengths drive the atoms into triple-occupation states, which dominate the pure ground state at these values of $U/J$ with $96.62\%$ triple-occupation (trimers) and $3.35\%$ double-occupation (dimers) contributions. In a lattice with nonvanishing disorder, the wave function therefore localizes solely around a small number of lattice sites. Such bunching is prohibited for bipartite settings with indistinguishable particles, where Pauli exclusion enforces a maximum of two atoms per site (see Sec.~\ref{sec:ML}), explaining the greater disorder susceptibility in the tripartite system. Decreasing the attraction strength diminishes the triplet bunching effect, leading to a less strong impact of disorder. This property comes at the cost of lower fidelity at very low disorder strengths, since the single-occupation and double-occupation probabilities rise accordingly.

At vanishing disorder, robust certification of four-dimensional tripartite entanglement is possible and for very strong disorder, $J\sigma_v\sim0.25$, two-dimensional tripartite entanglement can still be confidently certified. We thus witness multipartite entanglement for an extended disorder regime.
In the regime shown in Fig.~\ref{fig:fidelitytripartite}(b), the contribution from lattice disorder to bound tightness is less than, or of the same order of magnitude as, the contribution from state dephasing. However, the reduction in true fidelity through disorder dominates all other error sources considered.

\section{Generalization to Other Reference States}
\label{sec:repulsive}

Up to this point, we have shown the application of our method to ground states of Hubbard models with attractive interspecies interactions. We now develop generalizations to repulsive models of two or more atoms. As the entire process of measuring the $g_{\alpha\beta}$ coefficients is agnostic with regard to the measured state, the steps outlined in Eqs.~\eqref{eq:grpdecomposition}-\eqref{eq:projectionIntMC} can be applied in an identical manner, leaving the experimental procedure unchanged. However, the reference state $\Psi_{\mathrm{ref}}$ must change and therefore one must extract different coherences. The employed scheme for deriving fidelity lower bounds can, in principle, be applied to any reference state. However, the bound tightness, especially in the presence of dephasing noise, will generally depend on the properties of the chosen reference state, leaving room for optimization in a given experimental scenario.

\subsection{Two repulsively interacting atoms}

A suitable reference state for the ground state of a repulsive Hubbard model of two atoms in a lattice of $L$ sites may be given by an equal superposition of all nondimer states,
\begin{align}
    \ket{\Psi_\mathrm{ref}} = \frac{1}{\sqrt{L(L-1)}}\sum_{i\neq j}^L\ket{ij}, \label{eq:psi_ref}
\end{align}
which in turn means that the fidelity to that reference state is given by
\begin{align}
    F(\hat{\rho}, \Psi_\mathrm{ref}) = \frac{1}{L(L-1)}\sum_{\substack{i\neq j\\k\neq l}}^L\bra{ij}\hat{\rho}\ket{kl}.\label{eq:naive_fidelity}
\end{align}

The procedure to extract coherences by subtracting bounds on all other noncontributing coherences presented in Eq.~\eqref{eq:fidelbound} can then be adapted to remove coherences not of the type of Eq.~\eqref{eq:naive_fidelity}. This can be done without added complexity, as all coherences have the same weight in the fidelity and can be homogeneously extracted from contributing $g_{\alpha\beta}$ coefficients. This delivers a valid and accessible lower bound on $F(\hat{\rho}, \Psi_\mathrm{ref})$. However, since 
$\Psi_\mathrm{ref}$ is not natively given in a Schmidt-decomposed form, one first has to compute the Schmidt decomposition  $\ket{\Psi_\mathrm{ref}} = \sum_{i=1}^L \lambda_i \ket{\lambda_i}_\mathrm{A}\otimes\ket{\lambda_i}_\mathrm{B}$ with Schmidt coefficients $\lambda_i,~\lambda_1 \geq \lambda_2\ldots\geq\lambda_L$ in order to give the entanglement-dimension thresholds $\mathrm{B}_k(\Psi_\mathrm{ref}) = \sum_{i=1}^k\lambda_i^2$. In the case of $L=6$, we find $\lambda_1 = \sqrt{5/6}$ and $\lambda_2 = \ldots = \lambda_6 = \sqrt{1/30}$. This results in a high barrier of $\mathrm{B}_1 = 5/6$ to detect entanglement at all, while the higher thresholds are equally spaced between $\mathrm{B}_1$ and 1. This is, in essence, caused by the fact that our initial guess for a reference state is simply not that highly entangled, as can be seen through the entanglement entropy $S(\Psi_\mathrm{ref})=5/6\log(6/5)+1/6\log(30)\approx0.719$ compared to the MES used in the attractive case with $S(\Psi_\mathrm{MES})=\log(6)\approx1.792$. 

We can compensate this shortcoming by exploiting the additional structure of reference states of the form \eqref{eq:psi_ref}. As we show in Appendix~\ref{app:lambda1}, uniform nondimer reference states always have an associated Schmidt basis vector $\ket{\lambda_1}_\mathrm{A}\otimes\ket{\lambda_1}_\mathrm{B}=1/L\sum_{i,j=1}^L \ket{ij}$, an equal superposition of all states in the entire Hilbert space. If one now defines a new reference state $\ket{\Psi_\mathrm{ref}'}$ in terms of the same Schmidt basis but varies the value of $\lambda_1$ and uniformly adapts the remaining Schmidt coefficients to preserve normalization, one obtains a family of highly entangled states symmetric under lattice-site exchange. This is important, as the weight $w$ of coherences now only depends on whether they are of type dimer-dimer ($\bra{ii}\hat{\rho}\ket{jj}$, $w_d^d$), dimer-nondimer ($\bra{ii}\hat{\rho}\ket{jk}$, $w_d^{nd}$), or nondimer-nondimer ($\bra{ij}\hat{\rho}\ket{kl}$, $w_{nd}^{nd}\geq0$), independent of the specific lattice sites. This makes their extraction much simpler and the process more resilient against dephasing effects, as we show below. Explicit expressions for the weights $w$ are also given in Appendix~\ref{app:lambda1}. The fidelity for a generic reference state from that family then reads
\begin{align}
\begin{split}
    F(\hat{\rho}, \Psi_\mathrm{ref}') = w_{nd}^{nd} &\sum_{\substack{i\neq j\\k\neq l}}^L\bra{ij}\hat{\rho}\ket{kl} + w_d^d \sum_{\substack{i=1\\j=1}}^L  \bra{ii}\hat{\rho}\ket{jj}\\
    + w_d^{nd} &\sum_{\substack{i=1\\j\neq k}}^L \bra{ii}\hat{\rho}\ket{jk}+\bra{jk}\hat{\rho}\ket{ii}.\label{eq:fidelity_rep_nonuniform}
\end{split}  
\end{align}

The bound can then be derived as follows. First, one bounds the nondimer-nondimer contributions from below by taking the sum of all coefficients and subtracting the bounds of all other terms as originally shown in Eq.~\eqref{eq:fidelbound},
\begin{align}
    \begin{split}
        w_{nd}^{nd}\sum_{\substack{i\neq j\\k\neq l}}^L\bra{ij}\hat{\rho}\ket{kl}\geq&w_{nd}^{nd}\Biggl(\quad\sum\limits_{\mathclap{(\alpha,\beta)\in M}}\operatorname{Re}(g_{\alpha\beta})\\
         &\hspace{-2cm}-\sum\limits_{\mathclap{\substack{m,n,m',n' = 1\\m = n \lor m'=n'}}}^{L} \sqrt{\bra{m'n'}\hat{\rho}\ket{m'n'} \bra{mn}\hat{\rho}\ket{mn}}\Biggr) \eqqcolon \tilde{F}_{nd}^{nd}\label{eq:bound_nd_nd},
    \end{split}
\end{align}
where the second sum on the right-hand side includes dimer populations as well as all dimer coherences. Second, we have to bound the dimer-dimer terms. In principle, this can be done analogously to Eq.~\eqref{eq:bound_nd_nd} and would amount to subtracting bounds for all nondimer terms, which for a repulsive model are much larger than dimer-dimer terms. Even slight dephasing would cause a significant underestimation of these coherences and a corresponding loss of bound tightness. A more controlled approach is to bound the sum as,
\begin{align}
    \begin{split}
        &w_d^d \sum_{\substack{i=1\\j=1}}^L  \bra{ii}\hat{\rho}\ket{jj} = w_d^d\left(\sum_{i=1}^L \bra{ii}\hat{\rho}\ket{ii} + \sum_{\substack{i\neq j}}^L  \bra{ii}\hat{\rho}\ket{jj}\right)\\
        &\geq w_d^d\sum_{i=1}^L \bra{ii}\hat{\rho}\ket{ii} - \Bigl|w_d^d\Bigr|\sum_{\substack{i\neq j}}^L  \sqrt{\bra{ii}\hat{\rho}\ket{ii}\bra{jj}\hat{\rho}\ket{jj}}\eqqcolon \tilde{F}_{d}^{d}\label{eq:bound_d_d}
    \end{split}
\end{align}
where we have replaced the dimer-dimer coherences with the negative of their upper bound. This introduces a small bias in the case of $w_d^d\geq 0$, meaning that all states with nonvanishing dimer populations can no longer deliver a tight bound, but the loss of bound tightness from this term is largely independent of the level of dephasing and thus is much more stable. 

While fidelity bounds can be derived analogously for arbitrary reference states, this susceptibility to dephasing renders reference states with no structure in their entanglement spectrum less suitable in practice. Coherences would appear in the fidelity with widely varying weights and would require a large number of bounds in the style of Eq.~\eqref{eq:bound_nd_nd}. Dephasing leads to underestimation of many of those bounds as described above and ultimately to the loss of any usable bound.

A similar argument can be made for the the dimer-nondimer coherences as shown below:

\begin{align}
    \begin{split}
        w_d^{nd} &\sum_{\substack{i=1\\j\neq k}}^L \bra{ii}\hat{\rho}\ket{jk} + \bra{jk}\hat{\rho}\ket{ii}\geq\\
        -2\Bigl|w_d^{nd}\Bigr|&\sum_{\substack{i=1\\j\neq k}}^L \sqrt{\bra{ii}\hat{\rho}\ket{ii}\bra{jk}\hat{\rho}\ket{jk}}
        \eqqcolon\, \tilde{F}_{d}^{nd}.\label{eq:bound_d_nd}
    \end{split}
\end{align}
Combining all three partial bounds results in the measurable fidelity lower bound given by

\begin{align}
    F(\hat{\rho}, \Psi_\mathrm{ref}') \geq \tilde{F}_{nd}^{nd} + \tilde{F}_{d}^{nd} + \tilde{F}_{d}^{d} \eqqcolon\,\tilde{F}'.
\end{align}

\begin{figure}
    \centering
    \includegraphics[width=\columnwidth]{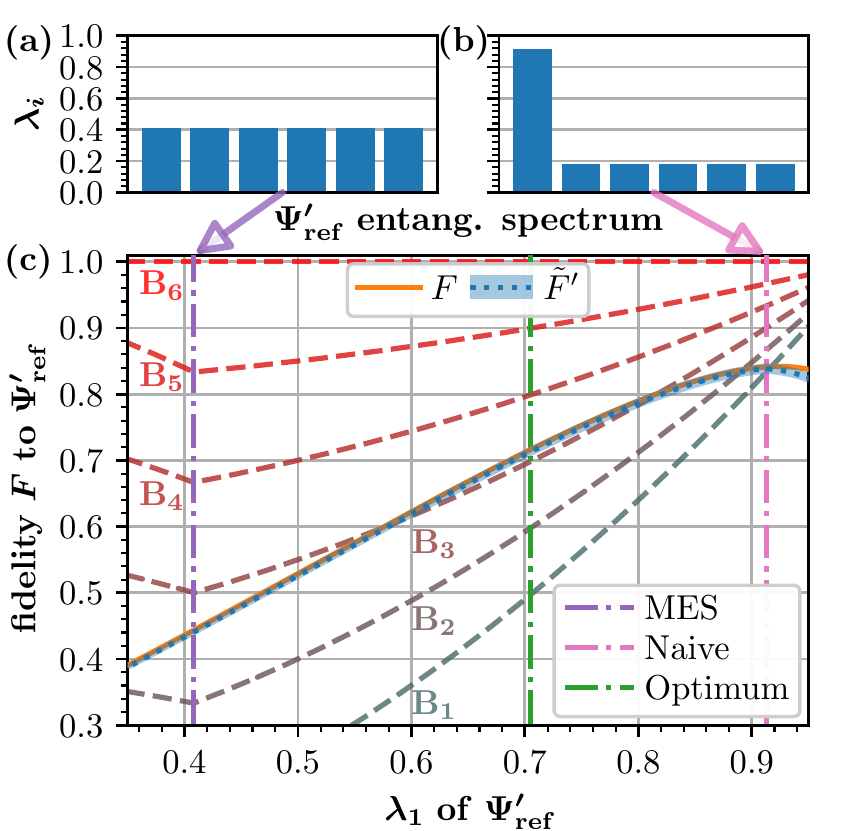}
    \caption{The entanglement spectra of the adapted MES with equally spaced $\mathrm{B}_k$ thresholds in \textbf{(a)} and of the reference-state choice with maximal fidelity in \textbf{(b)}. \textbf{(c)} By varying the first coefficient $\lambda_1$ of $\Psi_{\mathrm{ref}}'$ in postprocessing, we find the optimal $D_\mathrm{ent}=4$ at $\lambda_1\approx0.706$ for the ground state for the two atoms at \textbf{$U/J=30$}. The measurement was simulated using \num{5e4} shots in position and momentum space. The blue-shaded area represents the $1\sigma$ confidence interval of the bound.}
    \label{fig:optimization}
\end{figure}

Consequently, one can first perform the measurement scheme as originally introduced in Sec.~\ref{sec:bounds} and then optimize the certified entanglement dimension by varying the value of $\lambda_1$ of $\ket{\Psi'_\mathrm{ref}}$. The results of this optimization procedure for the two-atom ground state with $L=6$ are displayed in Fig.~\ref{fig:optimization}. The highly peaked entanglement spectrum of the ``naive" initial guess $\ket{\Psi_{\mathrm{ref}}}$ offers the best fidelity but insurmountable entanglement thresholds. On the other hand, a uniform spectrum, i.e., a maximally entangled state, delivers ideal bounds but at the cost of loss in fidelity. We find the highest certified entanglement dimension of $D_{\mathrm{ent}}=4$ at $\lambda_1\approx0.706$.

\subsection{Multiple atoms per species with repulsive interactions}

Next, we also investigate a repulsive system of $N$ atoms per spin state at half-filling ($L=2N$). Here, the adaptation of our method is much more straightforward than we saw before. In the attractive case, the Hubbard-model ground state was close to a superposition of states in which all atoms were bound in dimers, across all lattice-site combinations [cf. Eq.~\eqref{eq:MES_ML}]. Therefore, we used this MES as a reference state. In the repulsive regime, the ground state is close to a superposition of states with no dimers. At half-filling, for a given configuration of sites being occupied by species-A atoms, there is a unique configuration of species-B atoms realizing a dimer-free state, namely all B atoms occupying the sites not occupied by A atoms. This means that the standard choice of perfectly anticorrelated atom positions,
\begin{align}
    \begin{split}
        \ket{\Psi}_{\mathrm{ref}}^{N+N} &= \frac{1}{\sqrt{\binom{L}{N}}}\!\sum_{{\substack{m_i,n_i=1}}}^{L}\!\ket{m_1\dots m_N}_{\mathrm{A}} \otimes \ket{n_1\dots n_N}_{\mathrm{B}},\\
        &\mathrm{with}~m_i<m_{i+1},~n_i<n_{i+1},~m_i\neq n_j~\forall i,j,\label{eq:MES_repulsive}
    \end{split}
\end{align}
is also a MES, equivalent to Eq.~\eqref{eq:MES_ML} up to a permutation of the species-B basis states and thus ideally suited for our entanglement-detection scheme. One therefore only has to extract coherences of perfectly anticorrelated states instead of correlated ones, while leaving the rest of the scheme unaltered.

\begin{figure}
    \centering
    \includegraphics[width=\columnwidth]{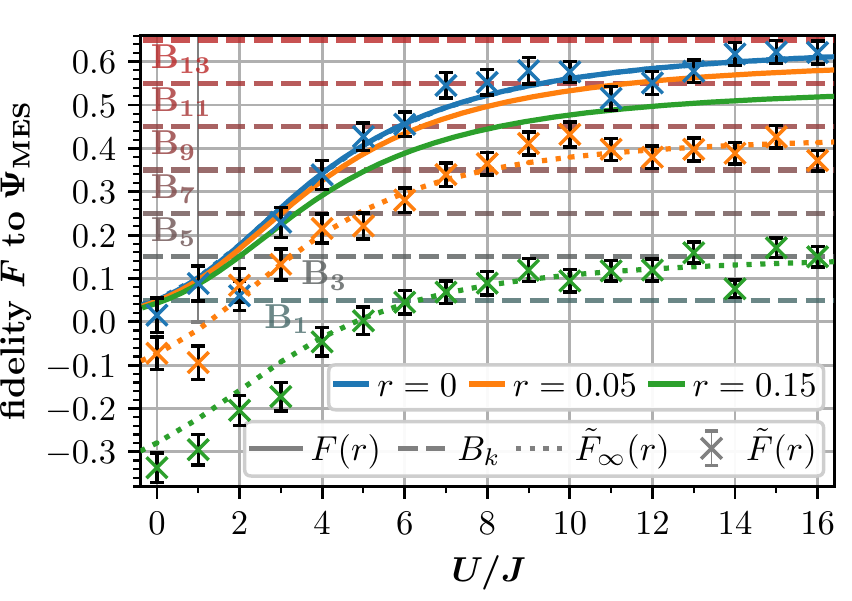}
    \caption{The numerical results for the entanglement-dimension certification of $3+3$ hard-core bosons in a lattice of size $L=6$ with repulsive interactions. The represented data mirror the above results for the same setting but with attractive interactions, in Fig.~\ref{fig:fidelityML}(b). The dotted lines again represent the infinite-measurement-statistics limit $\tilde{F}_{\infty}$ computed using exact coherences of $\hat{\rho}$.
    All measurements were simulated using \num{1e5} momentum-space and position-space samples each.}
    \label{fig:633_repulsive}
\end{figure}

We have simulated the extraction procedure for the ground state of a system of varying repulsive $U/J$ with $N=3$ hard-core bosons per species in a lattice with $L=6$. The results are shown in Fig.~\ref{fig:633_repulsive}.
Depending on the level of dephasing, certification of up to $D_\mathrm{ent}=13$ is feasible. Our data mirror previous data from our investigation for attractive systems in Fig.~\ref{fig:fidelityML}(b), where we have observed matching fidelities and bounds for exchanging $U\leftrightarrow-U$ and replacing the attractive MES in Eq.~\eqref{eq:MES_ML} with the repulsive MES from Eq.~\eqref{eq:MES_repulsive}, reminiscent of a particle-hole symmetry. This means that the method can be applied to Hubbard-model ground states over the entire range of interaction strengths, with particularly favorable properties in the half-filling case. But, also, multiatom scenarios away from half-filling can be treated. There, we do not have a unique particle-hole matching, but several contributions with nonvanishing configurations have to be considered in a reference state. For states close to half-filling the entanglement spectrum remains mostly flat but the lower the density in the lattice, the less informative is the knowledge of the positions of species A about species B. This causes a more strongly peaked entanglement spectrum, as we have observed in the case of two atoms on $L=6$ sites. To demonstrate that certification is still possible we performed, as an example, a simulated application of the method for the ground state of $N=2$ hard-core bosons per species in $L=6$ lattice sites at $U/J=-12$, using \num{5e4} samples in momentum and position space each. The naive uniform superposition of all nondimer states as $\ket{\Psi_\mathrm{ref}}$ yields a fidelity bound $\tilde{F}(\hat{\rho}, \Psi_\mathrm{ref})=0.682\pm0.021$. The entanglement spectrum of $\ket{\Psi_\mathrm{ref}}$ is given by $\lambda_1 = \sqrt{0.4}$, $\lambda_2=\ldots=\lambda_6=\sqrt{0.1}$, and $\lambda_7=\ldots=\lambda_{15} = \sqrt{1/90}$. This means that $\tilde{F}(\hat{\rho}, \Psi_\mathrm{ref})> \mathrm{B}_3 = \sum_{i=1}^3 \lambda_i^2 = 0.6$, so $D_\mathrm{ent}=4$ can be certified at $3\sigma$ confidence. Also, here one could conceive of a scheme to design better-suited reference states in the spirit of Fig.~\ref{fig:optimization}, which we leave for future investigations.

\section{Conclusions and Discussion}
\label{sec:conclusions}
\subsection{Summary}

We have presented a new method to bound the fidelity of few-body states of ultracold-atom systems to a highly entangled state. High fidelity indicates the presence of high-dimensional entanglement in the experimental state and can be used to bound entanglement quantifiers such as the entanglement dimension or the entanglement of formation. We have constructed lower bounds on the fidelity that are measurable in systems of ultracold atoms in optical lattices utilizing only position- and momentum-space measurements. A detailed study of the statistical significance and tightness of these bounds under realistic assumptions about experimental measurement conditions and noise sources indicates manageable experimental and statistical requirements. Interestingly, states that are highly mixed due to lattice-potential fluctuations retain their bound tightness to a high degree, allowing one to observe the disorder-induced reduction of ground-state entanglement. Generic white noise has been identified to cause linear decline of the tightness of our fidelity bound, while finite temperature has a comparably mild impact on bound tightness.
We have generalized this method to certify entanglement in multipartite systems and configurations with several atoms per spin state, requiring alterations to the coherence-extraction framework to account for partial indistinguishability. In these settings, we have demonstrated the feasibility of certifying up to $D_{\mathrm{ent}}=31$ entanglement dimensions for $4+4$ hard-core bosons and up to $D_{\mathrm{ent}}=4$ of genuine tripartite entanglement. Furthermore, by using reference states beyond the canonical MES, we have demonstrated the wide applicability of our method to quantum simulation experiments with itinerant particles in lattice geometries.

\subsection{Literature context}
\label{sec:literature}
Our work should be considered in the context of research lines focusing on efficient state tomography schemes or, leaving tomography out as an intermediate step, direct entanglement-detection methods. Here, we briefly review these research lines, commenting on their strengths and weaknesses compared to our method, without any claim of completeness. Full quantum state tomography in a bipartite system generally requires a number of different measurement bases that scale quadratically in the local Hilbert-space dimension \cite{parisQuantumStateEstimation2004}. This limits its applicability to very small system sizes \cite{Haffner2005}, despite significant advances in the efficiency of maximum-likelihood estimation \cite{Shang2017} and Bayesian tomography methods \cite{Granade2016}. A more economic scaling of experimental cost can be reached by restricting the state space in which the reference state is being searched. Examples for such approaches are compressed sensing tomography \cite{Gross2010, Kalev2015, Riofrio2017}, assuming that the prepared state has reduced rank, and methods using variational ansatz functions, such as neural-network quantum state tomography \cite{Torlai2018, Torlai2020, Carrasquilla2019, schmale_efficient_2022} or matrix-product state tomography \cite{Cramer2010, Baumgratz2013, Lanyon2017}, which restricts its search space to weakly entangled states—operating exactly in the opposite regime to the one targeted in this work. The drawback of this class of methods is that restricting the state space necessarily leads to bias, as it is generally not known whether the experimentally prepared state lies in the class of states representable by the ansatz.

For extracting properties of the entanglement spectrum, it is often not necessary to fully reconstruct the quantum state.
For example, if the global state of the system can be assumed to be pure, the entanglement spectrum can already be extracted from the state $\hat{\rho}_A$ of a subsystem. A variational approach to determine the entanglement Hamiltonian, i.e.,\ the logarithm of the reduced density matrix, has recently been demonstrated experimentally \cite{Kokail2021a, Kokail2021b,joshi_exploring_2023}.
Another notable approach is the use of random measurements to detect entanglement \cite{Elben2020} within the framework of shadow tomography \cite{Huang2020}. This framework can be applied to extract Schmidt-number witnesses by probing correlation matrices \cite{wyderka_probing_2023, liu_characterizing_2023}. However, the method requires large sample sizes and the implementation of Haar-random unitary operators, an open challenge for systems of itinerant particles.
If one is only interested in Rényi entanglement entropies, methods using multiple copies of the quantum state can be employed \cite{Islam2015}.
A related protocol uses ancillary particles to measure the entanglement spectrum directly for cold lattice-confined bosons \cite{pichler_measurement_2016}. While this method is very elegant, it poses stringent requirements on experimental capabilities.

The approach pursued in our work relies on measurable lower bounds on the fidelity to a highly entangled reference state for probing the entanglement dimension. A number of works have studied efficient methods for estimating fidelity, or at least bounding it, ranging from correlation-measurement-based approaches \cite{krenn_generation_2014, erker_quantifying_2017, Bavaresco2017} to variational methods \cite{Cerezo2020} and random Pauli-string measurements \cite{Flammia2011, DaSilva2011}. Often, these schemes are tailored to a specific experimental system, such as entangled photon pairs in the case of Refs.~\cite{erker_quantifying_2017, Bavaresco2017}, where the capability to measure in a pair of MUBs is exploited. This makes it difficult to apply these methods to other platforms, where these capabilities are not given. The strength of our proposal lies in the development of an entanglement-certification scheme that relies on techniques readily available to cold-atom experiments and is generally applicable to bipartite and multipartite scenarios realizable with this versatile quantum simulation platform.

\subsection{Scalability}
The term quantum simulation often entails the notion of scalability to system sizes that are beyond the reach of classical simulation methods, i.e., reaching the regime of quantum advantage.
Here, we summarize our findings on the scaling of both experimental and computational cost of our method as a function of lattice size $L$ and particle number per species $N$.

We have found an algebraic saturation of the SE of our fidelity bound for growing lattice sizes for $N=1$. Statistical requirements for faithful entanglement certification thus remain approximately constant for an extended regime of lattice sizes, which opens up one pathway to prepare highly entangled states in large lattices. However, increasing the number of atoms (keeping the density constant) in bipartite configurations results in an exponential increase of the SE, which consequently necessitates an increase of samples taken by nearly one order of magnitude to add an additional pair of atoms to the system. We note, however, that experimental sampling of momenta is achieved through fluorescence imaging, where all momenta of one atomic species are captured in a single image. Consequently, there is no inherent connection between the sampling rate and the system size. This allows comparably fast sample production in systems with several atoms compared to the creation of such samples by numerical simulations, where the computational complexity is linked to the number of atoms.

The data-processing routine used in this work consists of several steps: projection of the sampled momentum distribution onto modes of the momentum-basis expansion, basis change via formal matrix inversion of the matrix $\bm{Q}$ to correct for nonorthogonality, and coherence extraction. All these steps have exponential computational complexity scaling in $N$, which also prohibits application to genuine many-body systems. By contrast, the local Hilbert-space size, and thus the processing complexity, is only polynomial in the lattice size $L$. 

\subsection{Outlook}

The detection scheme proposed here can be generalized and extended in various ways. First, the momentum-space measurement, achieved by completely switching off the lattice potential, operates in the continuous domain. One could also envision only tuning the interparticle interaction strength to zero and allowing the particles of each species to undergo a tunneling evolution in the lattice before they are imaged. This would correspond to a measurement in a discrete basis complementary to the \textit{in situ} measurement, similarly giving access to coherences as the current scheme but avoiding the step of projecting measured data onto a function basis in continuous space. Furthermore, it would also relax the resolution requirements in momentum space, making the method accessible to a even broader range of contemporary experimental setups.
Second, entanglement-dimension witnesses based on measurements in two complementary bases may be developed analogously for other quantum simulation platforms. Examples are trapped ions, superconducting qubits, or Rydberg atoms, realizing spin systems, where the entangled subsystems consist of multiple spins, or qubits, with native local unitary transformations available to each specific platform. 
Furthermore, one could include a small number of additional measurement bases to give further constraints on state coherences, improving bound tightness, especially for higher atom numbers.
Finally, while investigating the impact of experimental imperfections on our ability to certify entanglement in realistic settings, we have found distinctly different signatures for pure state dephasing and lattice disorder. This implies that this bound could be used as a probe of disorder and localization in the prepared state. More generally, we would like to apply the developed method to more quantum states of interest, beyond Hubbard-model ground states, exploring the rich variety of entanglement phenomena accessible with cold-atom quantum simulators.

\section*{Acknowledgments}

We thank N.\ Friis, M.\ Huber, S.\ Jochim, P.\ Preiss, and G.\ Vitagliano for discussions and A.\ Braemer and M.\ Reh for valuable comments on the manuscript. We acknowledge support by the state of Baden-Württemberg through bwHPC (``High Performance Computing, Data Intensive Computing and Large Scale Scientific Data Management in Baden-Württemberg") and the German Research Foundation (DFG) through Grant No. INST 40/575-1 FUGG (JUSTUS 2  and HELIX compute clusters), Germany’s Excellence Strategy EXC2181/1-390900948 (the Heidelberg STRUCTURES Excellence Cluster), and within the Collaborative Research Center SFB1225 (ISOQUANT)—Project-ID 273811115.

\appendix
\numberwithin{equation}{section}

\section{Numerical methods}
\label{app:numerics}

All numerical results presented in this paper require a number of processing steps, ranging from synthetic sample generation to nonorthogonality corrections and coherence extraction, that come with computational complexity scaling exponentially with the size of the system. In the following we will briefly describe the numerical methods we used for optimizing the performance of classical processing steps, which have been crucial for reaching the largest reported system sizes.

The key technique for synthetic data generation is the sampling process for high-dimensional probability distributions. Computing the full distribution on a grid becomes prohibitively expensive, so a Monte Carlo type algorithm must be employed instead. Realizing that the momentum integral over Eq.~(\ref{eq:decomposition}) separates for each term, one can utilize a so called \textit{ancestral sampling} procedure \cite{Bishop2006}: Since the integrals can be split up, one can integrate out all but one of the momenta to obtain the marginal $p(k_1)$. After sampling $k_1$ from that distribution, one can fix $k_1$ and integrate out the rest, now leaving $k_2$ open to obtain the conditional probability distribution $p(k_2|k_1)$. This scheme can be repeated until all momenta are fixed and a complete sample is generated. Replacing $d$-dimensional integrals with the product of $d$ 1D integrals, that can be computed beforehand, greatly increases accessible system sizes.
The remaining 1-dimensional integrals $I(\delta)$ are of form
\begin{equation}
    I(\delta)=\int\mathrm{d}k\,|\tilde{w}(k)|^2\cos(d\delta k)\,.
\end{equation}
We find that $|\tilde{w}(k)|$ can be well approximated through a Gaussian $g(\mu=0,\sigma)$, which also agrees with experimental findings \cite{Bergschneider2019}. Replacing the Wannier envelope yields an expression of the form $I(\delta)\sim\exp(-(d\sigma\delta/2)^2)$. With this, the expression for $p(k_1)$ becomes a sum over terms weighted by the integral over the remaining $l-1$ momenta,  $I(\delta_2,\ldots,\delta_l) \sim \exp(-(d\sigma/2)^2\sum_{i=2}^l \delta_i^2)$. As these are exponentially small in $\sum_{i=2}^l \delta_i^2$, we can define a cutoff $\delta_c$ and neglect all terms for which $\sum_{i=2}^l \delta_i^2>\delta_c$.

A similar technique can be used to simplify the computation of the matrix elements of the basis overlap matrix $\bm{Q}$ [see Eq.~\eqref{eq:nonorthogonal}]. The matrix is needed to correct the measured coefficients $c_{\alpha\delta}$ for overlap with different nonorthogonal basis elements. In the two-atom case, each element is given by
\begin{equation}
    \begin{split}
        Q_{\alpha\beta}^{\tilde{\alpha}\tilde{\beta}}=\int&\mathrm{d}k_1\mathrm{d}k_2\,|\tilde{w}(k_1,k_2)|^4\\
        &\cos[d(\alpha k_1+\beta k_2)]\cos[d(\tilde{\alpha} k_1+\tilde{\beta} k_2)]\,.
    \end{split}
\end{equation}
By similar manipulations of the trigonometric functions under the integral, one can obtain the following factorized form 
\begin{equation}
    \begin{split}
        Q_{\alpha\beta}^{\tilde{\alpha}\tilde{\beta}}&=\frac{1}{2}\left(f(\alpha+\tilde{\alpha})f(\alpha-\tilde{\alpha})+f(\beta+\tilde{\beta})f(\beta-\tilde{\beta})\right)\\
        f(\gamma)&=\int\mathrm{d}k\,|\tilde{w}(k)|^4 \cos(\gamma k)\,.
    \end{split}
\end{equation}
Using the fact that $ f(\gamma) =  f(-\gamma)$, one only needs to evaluate $f(\gamma)$, where $\gamma \in \{0,\dots,2(L-1)\}$.

The formal inversion $\Vec{G} = \bm{Q}^{-1}\Vec{C}$ can be efficiently approached by exploiting that $\bm{Q}$ is hermitian and positive-definite and using Cholesky decomposition, $\bm{Q} = \bm{LL^\dagger}$ \cite{benoit_note_1924}, which gives the lower-diagonal matrix $\bm{L}$ acting as a preconditioner for $\bm{Q^{-1}}$. For large $\bm{Q}$, saving the dense $\bm{L}$ can become too costly, so that the iterative conjugate-gradient algorithm \cite{hestenes_methods_1952} becomes the more practical solution. Both algorithms are available within the {\scshape numpy} and {\scshape scipy} scientific computing libraries in {\scshape Python} \cite{harris2020array,2020SciPy-NMeth}, which we have used for our numerical simulations presented in this work.

\setcounter{equation}{0}
\begin{widetext}
    \section{Fitting parameters}
    \label{app:fit-parameters}
    In Table~\ref{tab:fit_params} we list the numerical fit models and fit parameters for all conducted fits appearing in this paper.
    \begin{table}[H]
        \centering
        \small
        \caption[Table of fitting parameters]{Table of fitting parameters \footnote{We replaced $J\sigma$ with $\sigma$ in disorder configurations for brevity of notation.}}
        \bgroup
        \def\arraystretch{1.25}%
        \begin{tabular}[c]{|>{\centering}m{0.05\textwidth}|>{\centering}m{0.20\textwidth}|>{\centering}m{0.18\textwidth}||>{\centering}m{0.17\textwidth}|>{\centering}m{0.17\textwidth}|>{\centering\arraybackslash}m{0.17\textwidth}|}
            \hline
            \multirow{2}{*}{Fig.}&\multirow{2}{*}{Description}&\multirow{2}{*}{Fitting Model}&\multicolumn{3}{c|}{Fitting Parameters}\\\cline{4-6}
            & & & a & b & c\\
            \hline\hline
            \ref{fig:statistics}(c) & SE $N_{\mathrm{s}}$ dependence& $\sigma_{\tilde{F}}(N_\mathrm{s})=a N_\mathrm{s}^{b}\vphantom{\Big\rangle}$ & $0.92^{+0.16}_{-0.13}$ & -0.48\tpm0.02 & -\\
            \hline
            \multirow{3}{*}{\ref{fig:dephasing}} & Dephasing  & \multirow{3}{*}{$F(r) = a r + b$} & & &\\
            &$F(r=0.00)$& &-0.76\tpm2e-16 & 0.787\tpm3e-17 & -\\
            &$\tilde{F}(r=0.00)$ & & -1.15\tpm0.03 & 0.788\tpm0.004 & -\\
            \hline
            \multirow{3}{*}{\ref{fig:lattice_disorder}} & Disorder  & \multirow{3}{*}{$\bar{F}(\sigma) = a e^{-b \sigma^c} +\frac{1}{L}$} & & &\\
            &$\bar{F}(r=0.00)$& & 5.6\tpm1.1 & 4.65\tpm0.19 & 0.256\tpm0.015\\
            &$\tilde{\bar{F}}(r=0.00)$ & & 3.7\tpm0.6 & 4.48\tpm0.13 & 0.311\tpm0.015\\
            \hline
            \multirow{6}{*}{\ref{fig:L-both-dependence}(a)} &  $L$ dependence order & \multirow{7}{*}{$F(L) = a L^{-b} + c$} & & &\\
            &$F(r=0.00)$ & & 0.902\tpm0.002 & 2.123\tpm0.003 & 0.76734\tpm6e-5\\
            &$F(r=0.05)$ & & 0.906\tpm0.002 & 2.115\tpm0.003 & 0.72998\tpm6e-5\\
            &$F(r=0.15)$ & & 0.916\tpm0.002 & 2.101\tpm0.003 & 0.65226\tpm5e5\\
            &$\tilde{F}(r=0.00)$ & & 0.95\tpm0.09 & 2.14\tpm0.14 & 0.766\tpm0.003\\
            &$\tilde{F}(r=0.05)$ & & 0.94\tpm0.06 & 1.94\tpm0.09 & 0.697\tpm0.003\\
            &$\tilde{F}(r=0.15)$ & & 0.93\tpm0.05 & 1.64\tpm0.08 & 0.565\tpm0.005\\
            \hline
              \multirow{7}{*}{\ref{fig:L-both-dependence}(b)} & $L$ dependence disorder &  & & &\\
            & $F(r=0.00)$ & \multirow{3}{*}{$F(L) = a e^{-b L}\phantom{~+ c}$} &0.9813\tpm0.0027 & 0.05350\tpm2.5e-4 & -\\
            &$F(r=0.05)$ & & 0.9347\tpm0.0025 & 0.05364\tpm2.5e-4 & -\\
            &$F(r=0.15)$ & & 0.8405\tpm0.0020 & 0.05387\tpm2.2e-4 & -\\
            \rule{0pt}{4ex}    
            &$\tilde{F}(r=0.00)$ & \multirow{3}{*}{$\tilde{F}(L) = a e^{-b L} + c$}& 1.040\tpm0.018 & 0.0534\tpm0.0026 &-0.047\tpm0.026\\
            &$\tilde{F}(r=0.05)$ & & 0.984\tpm0.013 & 0.0575\tpm0.0022 & -0.043\tpm0.019\\
            &$\tilde{F}(r=0.15)$ & & 0.910\tpm0.006 & 0.0644\tpm0.0017 & -0.070\tpm0.011\\
            \hline
            \ref{fig:L-resampling}(b) & SE $L$ dependence & $\sigma_{\tilde{F}}(L)=a L^{-b} + c\vphantom{\Big\rangle}$ & -0.0044\tpm1.6e-4 & 1.25\tpm0.06 & 0.00743\tpm2.4e-5\\
            \hline
            \multirow{3}{*}{\ref{fig:fidelityML}(c)} & $3+3$ disorder & \multirow{3}{*}{$\bar{F}(\sigma) = a e^{-b \sigma} + c$} & & &\\
            & $\bar{F}(r=0.00)$ & & 0.5197\tpm0.0012& 4.52\tpm0.05 & 0.0889\tpm9e-4\\
            & $\tilde{\bar{F}}(r=0.00)$ & & 0.604\tpm0.004 & 5.30\tpm0.06 & 0.0195\tpm9e-4\\
            \hline
            \ref{fig:finitesize} & SE $N$ dependence & $\sigma_{\tilde{F}}(N)=a e^{bN}\vphantom{\Big\rangle}$ & (4.4e-4\tpm5e-5) & 1.20\tpm0.04 & -\\
            \hline
            \multirow{3}{*}{\ref{fig:fidelitytripartite}(b)} & Tripartite disorder & \multirow{3}{*}{$\bar{F}(\sigma) = a e^{-b\sigma^c} +\frac{1}{L}$} & & &\\
            & $\bar{F}(r=0.00)$ & & 1.98\tpm0.22 & 6.50\tpm0.05 & 0.300\tpm0.013\\
            & $\tilde{\bar{F}}(r=0.00)$ & & 0.93\tpm0.04 & 8.58\tpm0.11 & 0.489\tpm0.011\\
            \hline
        \end{tabular}
        \egroup
        \label{tab:fit_params}
    \end{table}

\normalsize
\end{widetext}
\clearpage
\mbox{}
\clearpage

\section{Bound performance for thermal states}
\label{app:thermal}

\begin{figure}
    \centering
    \includegraphics{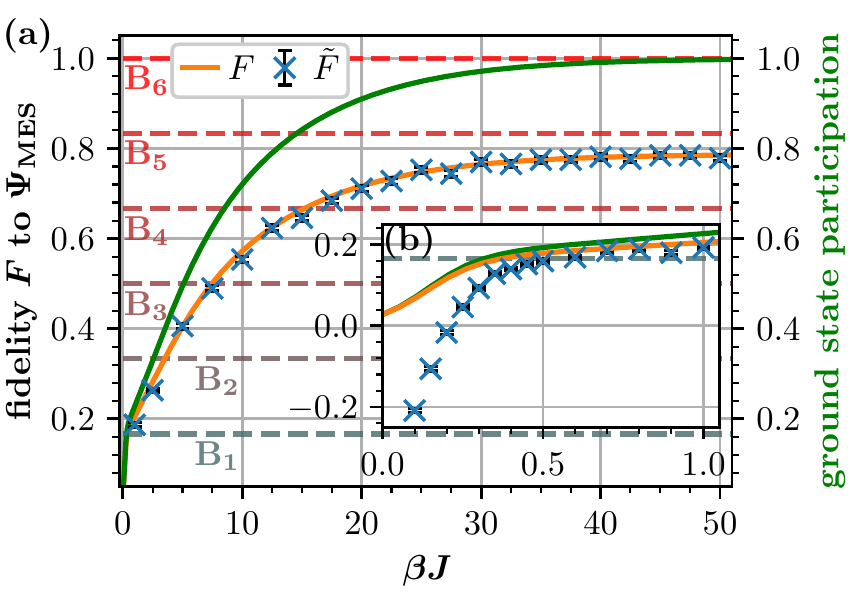}
    \caption{Numerical study of a thermal state of two atoms at $U/J = -12$ in a lattice with $L=6$ sites. (\textbf{a)} Fidelity $F(\hat{\rho}_T, \Psi_\mathrm{MES})$ in dependence of the normalized inverse ensemble temperature $\beta J$ on the left axis. On the right, ensemble ground-state participation amplitude. \textbf{(b)} Zoom in on values of $\beta J \leq 1$. The bound rapidly starts to loose tightness for $\beta J \lesssim 0.5$. As before, the data has been taken using \num{1e4} position-space and \num{2.5e4} momentum-space samples.}
    \label{fig:thermal}
\end{figure}

To ascertain the susceptibility of the derived fidelity bounds to thermal excitation, we have conducted additional simulations for thermal states, $\hat{\rho}_T\ \sim \exp(-\beta H)$,
of two attractively interacting distinguishable atoms in a lattice of size $L=6$. The resulting fidelities in dependence of the normalized inverse temperature, $\beta J$, are displayed in Fig.~\ref{fig:thermal}. In contrast to our results using white noise as a generic decoherence model, presented in Fig.~\ref{fig:dephasing}, we observe no significant loss of bound tightness for a broad temperature range of $\beta J \gtrsim 0.5$. This value of $\beta J$ translates to a ground-state fraction of $\approx19.3\%$ and an ensemble purity of $\approx0.164$. For even higher temperatures (smaller $\beta J$) the bound starts to deviate from $F$, as shown in the inset, but this is inconsequential for entanglement-dimension certification, as both $F$ and $\tilde{F}$ are lower than $\mathrm{B}_1$ and entanglement can no longer be witnessed.

These results indicate that the proposed bound is resilient against dissipation through coupling to a finite temperature bath. Thus, the method can be applied in an experimental setup where one cools directly into the Hubbard-model ground state, ending up at some finite temperature. We note that ground states may also be prepared by adiabatic deformations of the optical potential, starting with a localized dimer. Realistic modeling of the preparation process will depend on the concrete experimental setup and is left for future investigation.

\begin{widetext}
\section{Details on indistinguishable atom bipartite entanglement certification}
\label{app:indis}

Here, we want to give a full derivation of our fidelity bound for systems of multiple indistinguishable particles per species. We begin by giving the field operators $\hat{\Psi}_{\uparrow}(k),\hat{\Psi}_{\downarrow}(k)$, in terms of their lattice-site creation and annihilation operators, where we have introduced shortened notation $\hat{a}_{j} \coloneqq \hat{c}_{j,\uparrow}$ and $\hat{b}_{j} \coloneqq \hat{c}_{j,\downarrow}$. Operators within one bosonic (fermionic) species are subject to commutation (anticommutation) relations.  Both species are distinguishable via their spin degree of freedom, meaning that operators from different species always commute. Formally, these statements then read
    \begin{subequations}
        \begin{align}
            \begin{split}
                \hat{\Psi}_{\uparrow}(k)&=\omega(k)\sum_{j=1}^L e^{idkj}\hat{a}_j,\hspace{0.5cm}
                \hat{\Psi}_{\downarrow}(k)=\omega(k)\sum_{j=1}^L e^{idkj}\hat{b}_j,\label{eq:fieldops}
            \end{split}\\
            \left[\hat{a}^{\dagger}_i,\hat{a}_j\right]_\pm&=\left[\hat{b}^{\dagger}_i,\hat{b}_j\right]_\pm=\delta_{ij},\label{eq:comm_delta}\\
            \left[\hat{a}^{\phantom{\dagger}}_i,\hat{a}_j\right]_\pm&=\left[\hat{b}_i,\hat{b}_j\right]_\pm=\left[\hat{a}^{\dagger}_i,\hat{a}^{\dagger}_j\right]_\pm=\left[\hat{b}^{\dagger}_i,\hat{b}^{\dagger}_j\right]_\pm=0,\label{eq:comm_0}\\
            \left[\hat{a}^{\dagger}_i,\hat{b}_j\right]_-&=\left[\hat{b}^{\dagger}_i,\hat{a}_j\right]_- = \left[\hat{a}_i,\hat{b}_j\right]_- =\left[\hat{a}^{\dagger}_i,\hat{b}^{\dagger}_j\right]_- = 0,\label{eq:commuation}
        \end{align}
    \end{subequations}
    where we have used $[\cdot]_\pm$ to denote anticommutator and commutator, respectively, and $\tilde{\omega}(k)$ is the Fourier transform of the Wannier envelope.
    We can then rewrite the $2N$-atom momentum correlation function using the field operator form given in Eq.~\eqref{eq:fieldops} and obtain
    \begin{align}
        \begin{split}
            &\left\langle:\!\hat{\mathrm{n}}_{\uparrow}(k_1)\hat{\mathrm{n}}_{\uparrow}(k_2)\hat{\mathrm{n}}_{\downarrow}(k_3)\hat{\mathrm{n}}_{\downarrow}(k_4)\!:\right\rangle = \left\langle\hat{\Psi}_{\uparrow}^{\dagger}(k_1)\hat{\Psi}_{\uparrow}^{\dagger}(k_2)\hat{\Psi}_{\downarrow}^{\dagger}(k_3)\hat{\Psi}_{\downarrow}^{\dagger}(k_4)\hat{\Psi}^{\vphantom{\dagger}}_{\uparrow}(k_2)\hat{\Psi}^{\vphantom{\dagger}}_{\uparrow}(k_1)\hat{\Psi}^{\vphantom{\dagger}}_{\downarrow}(k_4)\hat{\Psi}^{\vphantom{\dagger}}_{\downarrow}(k_3)\right\rangle\\
            =&\Big|\underbrace{\tilde{\omega}(k_1)\tilde{\omega}(k_2)\tilde{\omega}(k_3)\tilde{\omega}(k_4)}_{\eqqcolon \tilde{\omega}(k_1,k_2,k_3,k_4)}\Big|^2\sum_{\substack{pp'qq'\\rr'ss'}}^L e^{-id[k_1(p-p')k_2(q-q')+k_3(r-r')+k_4(s-s')]}\left\langle\hat{a}^{\dagger}_p\hat{a}^{\dagger}_q\hat{b}^{\dagger}_r\hat{b}^{\dagger}_s\hat{a}^{\vphantom{\dagger}}_{q'}\hat{a}^{\vphantom{\dagger}}_{p'}\hat{b}^{\vphantom{\dagger}}_{s'}\hat{b}^{\vphantom{\dagger}}_{r'}\right\rangle\label{eq:pos_space_rep_ML}
        \end{split}
    \end{align}
    in analogy to the two-atom result in Eq.~\eqref{eq:pos_space_rep}. Note that we have used the normal ordered correlation function $\left\langle:\!\hat{\mathrm{n}}_{\uparrow}(k_1)\hat{\mathrm{n}}_{\uparrow}(k_2)\hat{\mathrm{n}}_{\downarrow}(k_3)\hat{\mathrm{n}}_{\downarrow}(k_4)\!:\right\rangle$, as it correctly represents single-atom-resolved measurements in momentum space \cite{preiss_high-contrast_2019, Folling2014}. This subtle distinction was not necessary for fully distinguishable particles, as the field operators $\hat{\Psi}_\sigma$ commute with each other. By inserting the coefficient expansion of $\hat{\rho}$ in the Fock basis and exploiting the relations in Eq.~\eqref{eq:commuation}, the expectation values in Eq.~\eqref{eq:pos_space_rep_ML} become
    \begin{subequations}
        \label{eq:expval}
        \begin{align}   \left\langle\hat{a}^{\dagger}_{p}\hat{a}^{\dagger}_{q}\hat{b}^{\dagger}_{r}\hat{b}^{\dagger}_{s}\hat{a}^{\vphantom{\dagger}}_{q'}\hat{a}^{\vphantom{\dagger}}_{p^{\prime}}\hat{b}^{\vphantom{\dagger}}_{s'}\hat{b}^{\vphantom{\dagger}}_{r'}\right\rangle &= \sum_{\substack{k<l,k'<l'\\m<n,m'<n'}}^L \hat{\rho}_{(klk'l')}^{(mnm'n')} \bra{klmn}\hat{a}^{\dagger}_p\hat{a}^{\dagger}_q\hat{b}^{\dagger}_r\hat{b}^{\dagger}_s\hat{a}^{\vphantom{\dagger}}_{q'}\hat{a}^{\vphantom{\dagger}}_{p'}\hat{b}^{\vphantom{\dagger}}_{s'}\hat{b}^{\vphantom{\dagger}}_{r'}\ket{k'l'm'n'}\\
            &=\sum_{\substack{k<l,k'<l'\\m<n,m'<n'}}^L \hat{\rho}_{(klk'l')}^{(mnm'n')} \bra{kl}\hat{a}^{\dagger}_p\hat{a}^{\dagger}_{q}\hat{a}^{\vphantom{\dagger}}_{q'}\hat{a}^{\vphantom{\dagger}}_{p'}\ket{k'l'}\bra{mn}\hat{b}^{\dagger}_r\hat{b}^{\dagger}_s\hat{b}^{\vphantom{\dagger}}_{s'}\hat{b}^{\vphantom{\dagger}}_{r'}\ket{m'n'},
        \end{align}
    \end{subequations}
    where we split the expectation value into a product between the two subsystems, or species. They can be evaluated independently using the commutation (anticommutation) relations from Eqs.~\eqref{eq:comm_delta}\&\eqref{eq:comm_0}, resulting in
    \begin{subequations}
        \label{eq:deltas}\begin{align}\bra{kl}\hat{a}^{\dagger}_p\hat{a}^{\dagger}_q\hat{a}^{\phantom{\dagger}}_{q'}\hat{a}^{\phantom{\dagger}}_{p'}\ket{k'l'} &= \bra{0}\hat{a}^{\vphantom{\dagger}}_{l}\hat{a}^{\vphantom{\dagger}}_{k}\hat{a}^{\dagger}_p\hat{a}^{\dagger}_q\hat{a}^{\vphantom{\dagger}}_{q'}\hat{a}^{\vphantom{\dagger}}_{p'}\hat{a}^{\dagger}_{k'}\hat{a}^{\dagger}_{l'}\ket{0} = (\delta_{p'k'}\delta_{q'l'}\mp\delta_{q'k'}\delta_{p'l'})(\delta_{pk}\delta_{ql}\mp\delta_{qk}\delta_{pl}),\\
            \bra{mn}\hat{b}^{\dagger}_r\hat{b}^{\dagger}_s\hat{b}^{\vphantom{\dagger}}_{s'}\hat{b}^{\vphantom{\dagger}}_{r'}\ket{m'n'} &= \bra{0}\hat{b}^{\vphantom{\dagger}}_{n}\hat{b}^{\vphantom{\dagger}}_{m}\hat{b}^{\dagger}_r\hat{b}^{\dagger}_s\hat{b}^{\vphantom{\dagger}}_{s'}\hat{b}^{\vphantom{\dagger}}_{r'}\hat{b}^{\dagger}_{m'}\hat{b}^{\dagger}_{n'}\ket{0} = (\delta_{r'm'}\delta_{s'n'}\mp\delta_{s'm'}\delta_{r'n'})(\delta_{rm}\delta_{sn}\mp\delta_{sm}\delta_{rn}).
        \end{align}
    \end{subequations}
    Inserting the above results from Eqs.~\eqref{eq:expval} and \eqref{eq:deltas} into Eq.~\eqref{eq:pos_space_rep_ML} gives a rather long expression. Therefore, for the moment, we leave out all terms connected solely to one of the two subsystems as shown in Eq.~\eqref{eq:subsystem} and reintegrate them later once all terms relating to the remaining subsystem have been sufficiently simplified.\\
    First, we resolve all open $\delta$ terms to eliminate $\{p,q,p',q'\}$ from the summation, which produces a sum of phase factors,
    \begin{subequations}
        \begin{align}
            &\sum_{\substack{k<l\\k'\!<l'}}^L\sum_{pp'qq'}^L \hat{\rho}_{(klk'l')}^{(mnm'n')}e^{-id[k_1(p-p')+k_2(q-q')]}\bra{kl}\hat{a}^{\dagger}_p\hat{a}^{\dagger}_q\hat{a}^{\phantom{\dagger}}_{q'}\hat{a}^{\phantom{\dagger}}_{p'}\ket{k'l'}\label{eq:subsystem}\\
           =&\sum_{\substack{k<l\\k'\!<l'}}^L \hat{\rho}_{(klk'l')}^{(mnm'n')}\left(e^{-id[k_1(k-k')+k_2(l-l')]}\mp e^{-id[k_1(k-l')+k_2(l-k')]}\mp e^{-id[k_1(l-k')+k_2(k-l')]}+e^{-id[k_1(l-l')+k_2(k-k')]}\right),\label{eq:nonbasis}
       \end{align}
   \end{subequations}
    each with a different permutation of site indices appearing the in bra and ket in Eq.~\eqref{eq:subsystem}.
    The sign of the different terms is determined by the underlying quantum statistics. Treating this sum of complex phase terms within the brackets in Eq.~\eqref{eq:nonbasis} (in combination with their complex conjugate counterparts) as the new basis functions would give the desired combinations of coherences as weights. However,  we find that these functions are linearly dependent, so that $\bm{Q}$ is generally rank deficient and thus cannot be inverted. Unambiguous reconstruction is therefore not possible. We overcome this issue by reorganizing terms. In a first step, we relabel all site indices such that all phases are of the same form, noted below the complex phases in Eq.~\eqref{eq:first_shuffle}. One has to take care to properly transport the conditions $k<l$ and $k'<l'$, which implement a second quantization picture, by introducing the corresponding Heaviside step functions $\theta(x)$ in Eq.~\eqref{eq:second_shuffle}. Using the symmetry (antisymmetry) of the density matrix $\hat{\rho}$ under particle exchange, one can rejoin all four terms into one term but without any restrictions regarding an ordering of $\{k, l\}$ or $\{k',l'\}$, as seen in Eq.~\eqref{eq:1stQ}. Finally, we replace two of the summation variables by the differences of the index pairs $\Delta k = k-k'$ and $\Delta l = l-l'$ to group together phases with the same factors appearing in the exponent [see Eq.~\eqref{eq:regroup}], 
   \begin{subequations}
    \begin{align}
        &\sum_{\substack{k<l\\k'\!<l'}}^L \hat{\rho}_{(klk'l')}^{(mnm'n')}\left(e^{-id[k_1(k-k')+k_2(l-l')]}\mp\underbrace{e^{-id[k_1(k-l')+k_2(l-k')]}}_{l'\longleftrightarrow\, k'}\mp\underbrace{e^{-id[k_1(l-k')+k_2(k-l')]}}_{l\,\longleftrightarrow\, k}+\underbrace{e^{-id[k_1(l-l')+k_2(k-k')]}}_{l'\longleftrightarrow\, k'\quad l\,\longleftrightarrow\, k}\right)\label{eq:first_shuffle}\\
        \begin{split}
               =&\sum_{\substack{kk'll'}}^L e^{-id[k_1(k-k')+k_2(l-l')]} \Biggl(\hat{\rho}_{(klk'l')}^{(mnm'n')}\theta(l-k)\theta(l'-k')\mp\underbrace{\hat{\rho}_{(kll'k')}^{(mnm'n')}}_{k'\,\longleftrightarrow\, l'}\theta(l-k)\theta(k'-l')\\
               &\hspace{3.85cm}\mp\underbrace{\hat{\rho}_{(lkk'l')}^{(mnm'n')}}_{k\longleftrightarrow\, l}\theta(k-l)\theta(l'-k')+\underbrace{\hat{\rho}_{(lkl'k')}^{(mnm'n')}}_{k'\longleftrightarrow\, l'~ k\,\longleftrightarrow\, l}\theta(k-l)\theta(k'-l')\Biggr)\label{eq:second_shuffle}
           \end{split}\\
           =&\sum_{\substack{kk'll'}}^L \hat{\rho}_{(klk'l')}^{(mnm'n')}e^{-id[k_1(k-k')+k_2(l-l')]}\label{eq:1stQ}\\
          = &\sum_{\substack{kl}}^L\sum_{\substack{\Delta k \coloneqq k-k'\\\Delta l \coloneqq l-l'\\\Delta k \leq \Delta l}} \hat{\rho}_{(klk'l')}^{(mnm'n')}\left(e^{-id[k_1\,\Delta k+k_2\,\Delta l]}+e^{-id[k_1\,\Delta l+k_2\,\Delta k)]}\right).\label{eq:regroup}
        \end{align}
    \end{subequations}
    Repeating above-described procedure for the second subsystem yields Eq.~\eqref{eq:final_ML}. This grouping of complex phases in combination with the Wannier envelope make up a complete basis and can thus be used to unambiguously reconstruct the corresponding basis weights,
        \begin{align}
        \label{eq:final_ML}
            \begin{split}
                &\langle:\!\hat{\mathrm{n}}_{\uparrow}(k_1)\hat{\mathrm{n}}_{\uparrow}(k_2)\hat{\mathrm{n}}_{\downarrow}(k_3)\hat{\mathrm{n}}_{\downarrow}(k_4)\!:\rangle=\big|\tilde{\omega}(k_1,k_2,k_3,k_4)\big|^2\sum_{\substack{kk'll'\\mm'nn'}}^L \hat{\rho}_{(klk'l')}^{(mnm'n')}e^{id[k_1(k-k')+k_2(l-l')+k_3(m-m')+k_4(n-n')]}\\
                &= \big|\tilde{\omega}(k_1,k_2,k_3,k_4)\big|^2 \sum_{\substack{klmn}}^L\sum_{\substack{\Delta k, \Delta l\\\Delta k \leq \Delta l}} \sum_{\substack{\Delta m, \Delta n\\\Delta m \leq \Delta n}}\hat{\rho}_{(kl(k-\Delta k)(l-\Delta l))}^{(mn(m-\Delta m)(n-\Delta n))}\\
                &\hspace{3.1cm}\left(e^{-id[k_1\,\Delta k+k_2\,\Delta l]}+e^{-id[k_1\,\Delta l+k_2\,\Delta k]}\right)\left(e^{-id[k_3\,\Delta m+k_4\,\Delta n]}+e^{-id[k_3\,\Delta n+k_4\,\Delta m]}\right).
            \end{split}
        \end{align}
     The summation over the differences $\Delta k$ and $\Delta l$ also allows for configuration where $k>l$, terms that were previously excluded in Eq.~\eqref{eq:first_shuffle}. This swapping of indices is equivalent to a particle exchange, which is accompanied by an additional minus sign for fermions in the corresponding matrix elements in $\hat{\rho}$. The alternating signs in Eq.~\eqref{eq:nonbasis} have effectively been shifted into the definition of the density matrix $\hat{\rho}$. This means that unlike the simple two-atom case, some coherences inherently acquire a sign here, which can lead to ``destructive interference" between coherences. The direct consequence is a loss in bound tightness for fermions, as observed in Sec.~\ref{sec:MLresults}.    
\end{widetext}
\clearpage
\mbox{}
\clearpage

\section{2+2 Atoms}
\label{app:2+2}

\begin{figure}[H]
    \centering
    \includegraphics[width=\columnwidth]{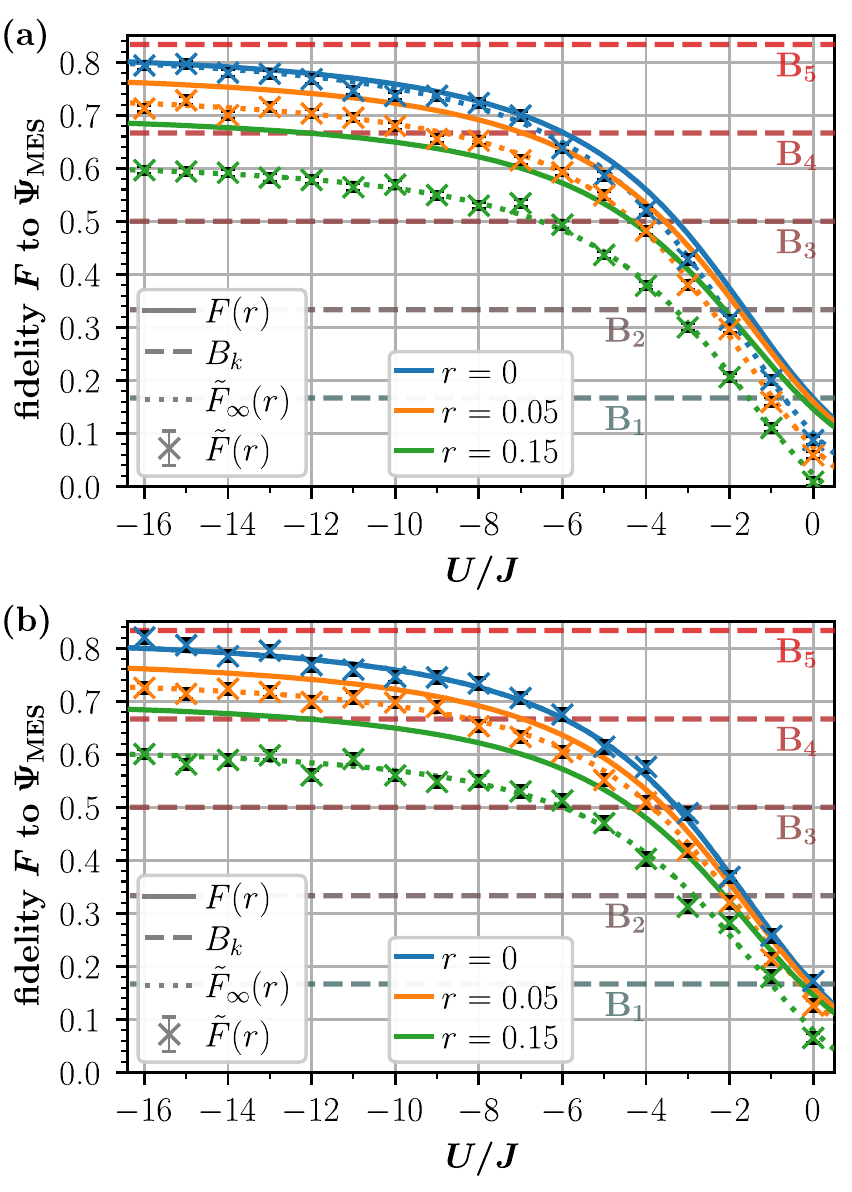}
    \caption{Numerical results for entanglement-dimension certification of $2+2$ indistinguishable atoms in a lattice with $L=4$. Dependence of the fidelity $F$ and the fidelity bound $\tilde{F}$ on the interaction-to-tunneling-strength ratio $U/J$ for pure and dephased states with \textbf{(a)} fermions and \textbf{(b)} hard-core bosons. The dotted line represents the infinite-measurement-statistics limit $\tilde{F}_{\infty}$ computed using exact coherences of $\hat{\rho}$. Differences between the two plots are most discernible for pure states at $U/J\sim0$. All measurements were simulated using \num{5e4} momentum-space and position-space samples each.}
    \label{fig:2+2}
\end{figure}

Figure \ref{fig:2+2} shows data for the case of $N=2$ atoms per species in analogy to Fig.~\ref{fig:fidelityML}. We observe that the loss of tightness for fermions is far less pronounced than in the case of $N=3$ atoms per species. The subtle differences between Fermi-Dirac and Bose-Einstein statistics can be most notably observed for very weak interactions strengths and pure states, where our bound underestimates the true state fidelity only for fermions.

\section{Tripartite entanglement-dimension bounds}
\label{app:bounds}

Here, we extend the concept of entanglement-dimension bounds $\mathrm{B}_k$  to tripartite reference states with generalized Schmidt decomposition, as given in Eq.~\eqref{eq:tripartite}, in close analogy to original work for bipartite states given in Refs.~\cite{guhne_detecting_2004,Fickler2014}. The general idea is again to give bounds on the maximal fidelity between some generalized reference state $\ket{\Psi}_L=\sum_{i=1}^L\lambda_i\ket{iii}$ and some state $\hat{\rho}_k$ with generalized Schmidt number $k$. This comparison can be made in a sensible way, as $\ket{\Psi}_L$ is already given in a form similar to the Schmidt decomposition of bipartite systems. All contributions are combinations of orthogonal basis states~$\ket{i}$ on the three subsystems and only appear once each. No unitary basis transformation therefore can reduce the number of states appearing in $\ket{\Psi}_L$, giving it the same role as the bipartite entanglement dimension. It is not necessary to  consider general mixed states, as convexity of the fidelity guaranties that fidelity is maximized through a pure state, so we restrict the proof to pure states only \cite{Fickler2014}. The highest possible fidelity between the reference and a pure state $\ket{\phi}_k=\sum_{lmn=1}^L c_{lmn}\ket{lmn}$ with $\ket{\phi}_k\in S_k$, where $S_k$ is the set of states with generalized Schmidt rank $k$, thus reduces to
\begin{align}
    \begin{split}
        &\sup_{\ket{\phi}_k\in S_k}F(\Psi_L,\phi_k)=\sup_{\ket{\phi}_k\in S_k}\bigl|\tensor[_L]{\braket{\Psi}{\phi}}{_{k}}\bigr|^2\\
    =&\sup_{\ket{\phi}_k\in S_k} \Biggl|\sum_{i=1}^L \lambda_ic_{iii}\Biggr|^2
    \end{split}
\end{align}
Without loss of generality, let $\lambda_1 \geq\lambda_2\ldots\geq\lambda_L$. Since $\ket{\phi}_k\in S_k$, at most $k$ of the $c_{iii}$ can take nonvanishing values. Additionally, wave-function normalization requires $\sum_i |c_{iii}|^2\leq1$. It is therefore clear that the supremum is realized with $c_{lmn} \neq 0$ only for $l = m = n \leq k$. Solving this optimization problem with a Lagrange multiplier, we arrive at $c_i = \lambda_i/\sqrt{\sum_{j=1}^k\lambda_j^2}$ for the optimal choice of coefficients. Inserting this into the fidelity yields
\begin{align}
    \sup_{\ket{\phi}_k\in S_k}F(\Psi_L,\phi_k)=\left|\frac{\sum_{i=1}^k\lambda_i^2}{\sqrt{\sum_{j=1}^k\lambda_j^2}}\right|^2 = \sum_{i=1}^k\lambda_i^2.
\end{align}

If one uses the generalized GHZ state $\ket{\mathrm{GHZ}}_L=1/\sqrt{L}\sum_i^L\ket{iii}$ as the reference state, one arrives at the same family of bounds as for the bipartite case,
\begin{align}
    \begin{split}
        &\sup_{\ket{\phi}_k\in S_k}F(\mathrm{GHZ}_L,\phi_k)= \frac{k}{L},
    \end{split}
\end{align}
as used in Sec.~\ref{sec:multipartite}. This resemblance is directly related to the restriction to Schmidt-decomposable states as reference states. Multipartite states in general cannot be brought into a form where each subsystem basis vector appears only once through some basis transformation. Therefore, this technique can never be expected to be able to detect all terms for a generic multipartite quantum state but, at most, the minimum of all local Hilbert-space dimensions. 

\section{Details on multipartite-entanglement certification}
\label{app:multipartite}

Extending the original scheme for entanglement certification to multipartite entanglement is straightforward but tedious. Here, we briefly want to give a starting point of how this extension is derived and present the final bound $\tilde{F}_\mathrm{coh}$. Like before, we decompose the momentum correlation function of three atoms $\langle\hat{\mathrm{n}}_{1}(k_1)\hat{\mathrm{n}}_{2}(k_2)\hat{\mathrm{n}}_{3}(k_3)\rangle$, in terms of coherences and consider phases picked up due to the Fourier transformation. This results in
 \begin{widetext}
    \begin{subequations}
        \begin{align}
            \begin{split}
                \langle\hat{\mathrm{n}}_{1}(k_1)\hat{\mathrm{n}}_{2}(k_2)\hat{\mathrm{n}}_{3}(k_3)\rangle &= \sum\limits_{\mathclap{\substack{a,b,c\,=\,1\\a',b',c' \,=\, 1}}}^L \phi_{a\ldots c'}(k_1,k_2,k_3)\bra{abc}\hat{\rho}\ket{a'b'c'},\label{eq:trimomcorr_phi}
            \end{split}\\
            \begin{split}
                \phi_{a\ldots c'}(k_1,k_2,k_3) &=|\tilde{\omega}(k_1,k_2,k_3)|^2 \exp\bigl\{-id[(a-a')k_1 + (b-b')k_2 + (c-c')k_3]\bigr\}.\label{eq:triphi}
            \end{split}
        \end{align}
    \end{subequations}
 \end{widetext}
We label the three distinguishable atom species $\{1,2,3\}$, with their respective lattice-site indices $\{a,b,c\}$ and $\{a',b',c'\}$ for the bra and ket states. This description can be expressed analogously to Eqs.~\eqref{eq:grpdecomposition} in trigonometric basis functions of all three lattice momenta $k_1,k_2,$ and $k_3$. Special care has to be taken to avoid double counting by adapting the set $M$ of of admissible lattice gap sets to again enforce
\begin{align}
  (\alpha,\beta,\gamma)  \in M \Rightarrow(-\alpha,-\beta,-\gamma)\notin M \lor  (\alpha,\beta,\gamma) = (0,0,0).    
\end{align}
 This is necessary to be able to do the full reconstruction of the momentum correlation function, since the true coefficients $g_{\alpha\beta\gamma}$ have to be obtained from the full distribution of measured coefficients $c_{\alpha\beta\gamma}$ first. The redefined set $M$ for three atomic species is given in Eq.~\eqref{eq:triM}. All remaining steps outlined in Eqs.~\eqref{eq:projectionInt} to \eqref{eq:CSI} can be adapted analogously, such that one arrives at the final result for the bound of the coherence contributions as follows:
\begin{widetext}
\begin{subequations}
    \begin{align}
        \begin{split}
           \tilde{F}_{\mathrm{coh}}(\hat{\rho}, \Psi_{\mathrm{MES}}) &= \sum\limits_{\mathclap{\delta=1}}^{L-1}\left(\frac{\operatorname{Re}(g_{\delta\delta\delta})}{L}-2\sum\limits_{\mathclap{\substack{a,b,c = 1\\a\neq b \lor b\neq c}}}^{L-\delta} \frac{\sqrt{\bra{a'b'c'}\hat{\rho}\ket{a'b'c'} \bra{abc}\hat{\rho}\ket{abc}}}{L}\right)\label{eq:trifidelbound}\\
           &\hspace{2cm}\mathrm{with}\quad a' \coloneqq a + \delta \quad b' \coloneqq b + \delta \quad c' \coloneqq c + \delta,
        \end{split}\\
        \begin{split}
            g_{\alpha\beta\gamma} &= 2\sum\limits_{a,b,c=1}^{L}\bra{abc}\hat{\rho}\ket{(a+\alpha), (b+\beta), (c+\gamma)}\quad \mathrm{with}\quad a+\alpha, b+\beta, c+\gamma\in\{1\ldots L\},\quad g_{000} = 1
            \label{eq:tricoeffs}
        \end{split}\\
        \begin{split}
            M = \Bigl\{(\alpha,\beta,\gamma&) \in \{-(L-1),\ldots,L-1\}^3 ~\Big|~\alpha\geq0 \land(\beta\geq0\lor\alpha>0) \land(\gamma\geq0\lor\beta>0\lor\alpha>0)\Bigr\}.\label{eq:triM}
        \end{split}
    \end{align}
\end{subequations}
\end{widetext}
\section{Schmidt basis properties of anticorrelated reference states}
\label{app:lambda1}

Here, we show some generic properties of the Schmidt decomposition of potential reference states $\ket{\Psi_\mathrm{ref}}$, which are exploited in the main text for optimizing Schmidt dimension witnesses for Hubbard-model ground states in the repulsive regime. Regarding a two-atom configuration in a lattice with $L$ sites, an intuitive choice is the uniform superposition of all nondimer states,
\begin{align}
    \ket{\Psi_\mathrm{ref}} = \frac{1}{\sqrt{L(L-1)}}\sum_{i\neq j}^L\ket{ij},
\end{align}
on which we need to perform a Schmidt decomposition to compute the entanglement-dimension bounds.
The first step consists of a singular value decomposition of the wave-function-coefficient matrix $\bm{C}$ defined through 
\begin{align}
    \bm{C}_{ij} = \braket{ij}{\Psi_\mathrm{ref}}=
    \begin{cases}
    0 & \mathrm{for}~i=j\\
    \frac{1}{\sqrt{L(L-1)}} & \mathrm{for} ~ i\neq j
    \end{cases}.
\end{align}
This matrix is symmetric and thus the absolute value of its eigenvalues are equal to its singular values. It is clear that such a matrix always has an eigenvector $\ket{\lambda_1}_\mathrm{A}=\ket{\lambda_1}_\mathrm{B}=1/\sqrt{L}\sum_{i=1}^L\ket{i}$, since every row of $\bm{C}$ contains the same number, $L-1$, of constant coefficients. This yields the eigenvalue of $\lambda_1 = (L-1)/\sqrt{L(L-1)} = \sqrt{(L-1)/L}$ and a corresponding Schmidt vector of 
\begin{equation}
    \ket{\lambda_1} = \ket{\lambda_1}_\mathrm{A}\otimes\ket{\lambda_1}_\mathrm{B} = \frac{1}{L}\sum_{i,j=1}^L\ket{ij}\label{eq:lambda_1}.
\end{equation}
Consequently, we can split up our reference state as
\begin{equation}
    \ket{\Psi_\mathrm{ref}} = \lambda_1 \ket{\lambda_1} + \sum_{i=2}^L\lambda_i\ket{\lambda_i} \eqqcolon \lambda_1 \ket{\lambda_1} + \lambda'\ket{\lambda'},
\end{equation}
where we have included all remaining contributions in a normalized state,
\begin{equation}
    \begin{split}
        \ket{\lambda'} =& \left(\sqrt{\frac{1}{L-1}} - \sqrt{\frac{L-1}{L^2}}\right)\sum_{i\neq j}^L\ket{ij}\\
        &-\sqrt{\frac{L-1}{L^2}}\sum_{i=1}^L\ket{ii},\label{eq:lamba_prime}
    \end{split}
\end{equation}
and $\lambda'=1/\sqrt{L}$.
 Note that both $\ket{\lambda_1}$ and $\ket{\lambda'}$ are symmetric under lattice-site exchange, such that any superposition of these states will have the same symmetry. This means that one can define the one-parameter family of reference states 
 \begin{equation}
     \ket{\Psi_\mathrm{ref}'}(\lambda) = \lambda \ket{\lambda_1} + \sqrt{1-\lambda^2} \ket{\lambda'}
 \end{equation}
 by varying the relative weight between them.

Finally, we compute the weights $w$ of coherences in the fidelity, introduced in Eq.~\eqref{eq:fidelity_rep_nonuniform}, as a function of the parameter $\lambda$. Using the above-discussed symmetry, we know that all dimer-dimer terms $\bra{ii}\hat{\rho}\ket{jj}$ must contribute equally, giving them a shared weight $w_d^d$. Analogous arguments can be made for dimer-nondimer coherences $\bra{ii}\hat{\rho}\ket{jk}$ with weight $w_d^{nd}$ and finally with nondimer-nondimer contributions $\bra{ij}\hat{\rho}\ket{kl}$ with weight $w_{nd}^{nd}$. These weights are given by
\begin{equation}
\begin{split}
        w_k^l = \ketbra{\Psi_\mathrm{ref}'}_k^l =& \lambda^2\ketbra{\lambda_1}_k^l + (1-\lambda^2)\ketbra{\lambda'}_k^l\\
        +& \lambda\sqrt{1-\lambda^2}\left(\ketbra{\lambda_1}{\lambda'}_k^l+\ketbra{\lambda'}{\lambda_1}_k^l\right),
\end{split}
\end{equation}
where the notation $\ketbra{\cdot}_k^l$ refers to the matrix element of the projector with respect to some basis elements $k$ and $l$, placeholders for dimer or nondimer states.
By plugging in the definitions of the two states from Eq.~\eqref{eq:lambda_1} and \eqref{eq:lamba_prime}, we obtain the following expressions for the three different weights as follows,
\begin{subequations}
    \begin{align}
        w_{nd}^{nd} &= \frac{1+\lambda^2(L-2)}{L^2(L-1)}+\frac{2\lambda\sqrt{1-\lambda^2}}{L^2\sqrt{L-1}}\label{eq:w_nd_nd},\\
        w_{d}^{nd} &= \frac{2\lambda^2-1}{L^2}-\frac{(L+2)\lambda\sqrt{1-\lambda^2}}{L^2\sqrt{L-1}},\\
        w_{d}^{d} &= \frac{L-1 - \lambda^2(L-2)-2\lambda\sqrt{1-\lambda^2}\sqrt{L-1}}{L^2}.
    \end{align}
\end{subequations}
It is clear from Eq.~\eqref{eq:w_nd_nd} that, for all $L\geq2$, one has $w_{nd}^{nd}\geq0$, since $\lambda\in[0,1]$.
\bibliography{mybib}

\end{document}